\documentclass[11pt,a4paper,nofootinbib,showpacs]{article}
\pdfoutput=1

\usepackage{jcappub}

\usepackage[utf8]{inputenc}
\usepackage[english]{babel}
\usepackage[T1]{fontenc}
\usepackage{graphicx}
\usepackage{amsmath}
\usepackage{subfig}
\usepackage{hyperref}
\usepackage{color}
\usepackage{amssymb}
\usepackage[font=scriptsize,labelfont=bf]{caption}
\usepackage{appendix}
\usepackage{slashed}
\usepackage{slantsc}
\usepackage{braket}
\usepackage{mathrsfs}
\usepackage{subfig}
\usepackage{placeins}
\usepackage{fullpage}
\usepackage{sidecap}
\usepackage{float}
\usepackage{wrapfig}
\usepackage[dvipsnames]{xcolor}
\usepackage{dsfont}
\usepackage{cleveref}
\usepackage{times}
\usepackage{natbib}
\usepackage{epsfig}
\usepackage{bm}
\usepackage{amsmath}
\usepackage{amssymb}

\newcommand{\aj}{Astron. J.}
\newcommand{\aap}{Astron. Astrophys.}
\newcommand{\cqg}{Class. Quant. Grav.}
\newcommand{\mnras}{Mon. Not. R. Astron. Soc.}

\newcommand{\jcap}{J. Cosmol. Astropart. Phys.}

\newcommand{\apj}{Astrophys. J.}
\newcommand{\prd}{Phys. Rev. D}

\usepackage{color}
\definecolor{red}{rgb}{1.00,0.00,0.00}

\begin{document}

\title{Correlation function of the luminosity distances}
\author[a,c]{Sang Gyu Biern}
\author[a,b]{and Jaiyul Yoo}
\affiliation[a]{Center for Theoretical Astrophysics and Cosmology,
Institute for Computational Science, University of Z\"urich,
Winterthurerstrasse 190, CH-8057, Z\"urich, Switzerland}
\affiliation[b]{Physics Institute, University of Z\"urich,
Winterthurerstrasse 190, CH-8057, Z\"urich, Switzerland}
\affiliation[c]{Asia Pacific Center for Theoretical Physics, Pohang, 790-784, Korea}
\emailAdd{sgbiern@physik.uzh.ch}
\emailAdd{jyoo@physik.uzh.ch}

\abstract{ We present the correlation function of the luminosity distances in a flat $\Lambda$CDM universe. Decomposing the luminosity distance fluctuation into the velocity, the gravitational potential, and the lensing contributions in linear perturbation theory,  we study their individual contributions to the correlation function. The lensing contribution is important at large redshift ($z\gtrsim 0.5$) but only for small angular separation ($\theta \lesssim 3^\circ$), while the velocity contribution dominates over the other contributions at low redshift or at larger separation.  However, the gravitational potential contribution is always subdominant at all scale, if the correct gauge-invariant expression is used. The correlation function of the luminosity distances depends significantly on the matter content, especially for the lensing contribution, thus providing  a novel tool of estimating cosmological parameters. }



\maketitle
\flushbottom
 
\section{Introduction}

 
 One of great mysteries of current physics is the accelerating expansion of the universe. This acceleration was first revealed by measuring the luminosity distances of distant supernovae~\cite{RIFIET98,PEADET99}, and following measurements of the microwave background  anisotropies~(e.g. \cite{PLANCK13}) and the acoustic peak position in galaxy clustering~(e.g. \cite{EIZEET05}) have supported this phenomenon.  However, despite the fact that the universe is inhomogeneous and anisotropic, the effects of inhomogeneity have been commonly dismissed in interpreting the luminosity distance measurements. 
 
 There are several theoretical efforts of deriving  the luminosity distance fluctuation, accounting for the inhomogeneities in our Universe~\cite{SASAK87,BODUGA06,HUGR06,YOSC16,BIYO16}. The fluctuation in the luminosity distance  was first derived by Sasaki~\cite{SASAK87} from the Sachs approach. Later studies computed the fluctuation in the luminosity distance  by utilizing the Jacobi mapping~\cite{BODUGA06}, the Geodesic-Light-Cone gauge approach~\cite{BEGAET12a,BEMAET12a,BEGAET13,BEDUET14,FAGAET13,GAMAET11}, and the geometric approach~\cite{JESCHI12,YOO14a}. 
 
 
 
    Using the full expression of the luminosity distance, the angular  power spectrum was derived in~\cite{BODUGA06}. The detectability of the power spectrum and the correlation function of the gravitational lensing contribution was investigated in~\cite{COHOHU06,SNMB16}.  These studies show that the lensing power spectrum/correlation can be measured in a high-$z$ and a deep survey such as the Large Synoptic Sky Telescope survey (LSST) ~\cite{LSST},  and  by using the supernova lensing dispersion, \cite{BERY2015,BE14,BEKA14} present the way of constraining the properties of the primordial power spectrum.   In addition to the lensing contribution, there exist other contributions to the luminosity distance, such as the velocity and the gravitational potential, and these contributions are all correlated. Thus, a proper computation of its correlation function should account for all these additional components. In particular, the gravitational potential contribution to the luminosity distance correlation has not been {\it properly} considered in literature.
    
    
    Furthermore, the correlated fluctuations in the luminosity distance can significantly affect the covariance matrix of the (apparent-) magnitude of supernovae, used for the cosmological parameter estimation in supernova surveys.  In~\cite{HUGR06}, it was shown that the systematic errors from the velocity contribution, induced by the large scale structure of the universe, can substantially bias the covariance matrix of the magnitude of supernovae. This study shows that there is a significant degradation in the determination of the dark energy equation of state.     
    
        
    

    Within the framework of cosmological perturbation theory, the observed luminosity distance fluctuation at the linear order consists of the velocity at the positions of observation and source, the lensing, and the gravitational potential contributions, but either the velocity at the observer position and/or the gravitational potential contributions  are not considered in the previous luminosity distance correlation related studies~(e.g.,~\cite{BEDUET14,MAC16}).   First, although the expressions of all the contributions are presented in literature, the gravitational potential contribution is always neglected in numerical computations related to the luminosity distance correlation function, arguing that it is suppressed compared to the other contributions. Second, the contribution of the velocity at the observation is turned off by hand because the observer's peculiar motion is corrected in the luminosity distance measurement by using the dipole of the cosmic microwave background radiation (CMB) observation.
    
    In~\cite{BIYO16}, it was demonstrated that the commonly used expression for the luminosity distance often lacks the coordinate lapse $\delta\tau_o$ at the observer position and its variance is infrared divergent. This infrared divergence in the luminosity distance variance originates from the gravitational potential contribution, and hence one naturally expects that the gravitational potential contribution can be the most dominant contribution in the expression of the previous studies. Despite the presence of such infrared divergences,  this pathology in the luminosity distance has been commonly dismissed  by considering only the sub-horizon perturbation mode (i.e., by introducing an {\it ad hoc} cutoff $k_\text{IR}$ as $k_\text{IR}\simeq \mathcal H_o$, where $\mathcal H_o$ is the current Hubble parameter). In fact, as shown in~\cite{BIYO16}, the luminosity distance is devoid of the infrared divergence when one considers the missing coordinate lapse and all the other contributions of the gravitational potential to the luminosity distance. Here we compute the gravitational potential contribution to the correlation function, and we show that it always contributes little to the correlation function of the luminosity distances.

    Furthermore, we also consider the velocity contribution at the observation position for two physical reasons: gauge invariance and equivalence principle. Without the velocity contribution at the observer position, the gauge-invariance of the expression for the luminosity distance is broken, and the uniform gravity-mode affects the resulting outcome, in direct conflict with the equivalence principle, as shown in~\cite{BIYO16,JESCHI12}. Practically, as opposed to the gravitational potential contribution, the velocity contribution at the observation is substantial, especially at small redshift. Thus, neglecting the  velocity contribution at the observation  causes not only the theoretical problems but also significant numerical errors. Therefore, we consider all the contributions to the luminosity distance correlation function. We further investigate how the  correlation function varies with respect to the matter content. 

   Before we proceed, we stress the critical difference in various terms
referring to the peculiar velocity at the observer position. 
Within the framework of cosmological perturbation theory, the peculiar velocity field is well defined, and the  peculiar velocity at the observer position is naturally called the peculiar velocity of the observer. However, these perturbations are gauge-dependent; their values change depending on our choice of coordinate systems. For example, the peculiar velocity field in the synchronous gauge is zero everywhere at the linear order in perturbations, while it is  non-vanishing in the conformal Newtonian gauge. Therefore, these peculiar velocities at the observer position also depend on gauge choice, and these perturbations themselves cannot have any physical meaning.

However, there exists a physically well-defined peculiar velocity of the
observer often used in literature, with which we can transform to a
frame where the CMB dipole vanishes. This peculiar velocity ($v_\text{dip}\simeq
371~\text{km/s}$~\cite{Planck:dipole}) is independent of our gauge choice, and it is $not$ identical to the peculiar velocity at the observer position in either the conformal Newtonian gauge or the synchronous gauge. It will be  referred to as the peculiar velocity of the observer with respect to the CMB rest frame, or the observed peculiar velocity. We will discuss its impact on the luminosity distance measurements in Sec.~\ref{Sec:discussion}. Unless mentioned explicitly, the peculiar velocity in this paper means the peculiar velocity in cosmological perturbation theory.

 This paper is organized as follows: In section \ref{Sec:review}, we review the linear-order luminosity distance fluctuation and decompose it into the velocity, the gravitational potential, and the lensing contributions. In section \ref{Sec:correlation}, we present the analytic expressions of the correlation functions of the velocity, the lensing, and the (simplified) gravitational potential contributions. The numerical results are shown in section \ref{Sec:results}:  the angular correlation function in section \ref{Sec:angular_correlation}, and the three-dimensional two-point correlation function in section \ref{Sec:2pt_correlation}. Especially, in section \ref{Sec:angular_correlation}, we investigate the angular correlation function of the luminosity distances with different cosmological parameter sets. In section \ref{Sec:discussion}, we summarize and discuss our new results. In appendix~\ref{App:GP}, we present the explicit expression of the auto-correlation function of the gravitational potential contribution.
 

%

\section{Luminosity distance in an inhomogeneous universe}
\label{Sec:review}
In this section we briefly summarize the linear-order calculations of the luminosity distance. The fluctuation $\delta\mathcal D_L$ in the luminosity distance $\mathcal D_L$ is given by~\cite{YOO14a}:
\begin{equation}
\label{Eq:DL}
	\mathcal D_L\equiv \bar{\mathcal D}_L(z)\left(1+\delta\mathcal D_L\right),~~~~~~~~~\delta\mathcal D_L=\delta z+\frac{\delta r}{\bar r_z}+\Xi-\kappa\, ,
\end{equation}
where $\bar{\mathcal D}_L(z)$ is the luminosity distance in a homogeneous universe, $\delta z$ is the redshift distortion, $\delta r$ is the radial distortion, $\Xi$ is the frame distortion, and $\kappa$ is the angular distortion.  

Thanks to the gauge-invariant calculation of the luminosity distance in~\cite{SASAK87}, we can obtain a consistent result in any gauge choice. From now on, we shall study the luminosity distance fluctuation in the conformal Newtonian gauge. In this gauge, we decompose the luminosity distance fluctuation into the velocity $\delta\mathcal D_L^V$, the lensing $\delta\mathcal D_L^\text{lens.}$, and the gravitational potential $\delta\mathcal D_L^\Psi$ contributions i.e., $\delta\mathcal D_L = \delta \mathcal D_L^V+\delta\mathcal D_L^\text{lens.}+\delta\mathcal D_L^{\Psi}$.\footnote{In our previous paper \cite{BIYO16}, we express the gravitational potential contribution without the coordinate lapse $\delta\tau_o$ at the observation. However, in this paper we define the gravitational potential contribution by including the contribution of $\delta\tau_o$. The term $\delta\mathcal D_L^{\Psi+\delta\tau_o}$ in~\cite{BIYO16} corresponds to the gravitational potential contribution in this paper.} The velocity (lensing) contribution contains   single (double) spatial derivative of the gravitational potential $\Psi$ i.e., $V_\| \propto \partial_\|\Psi$ ($\Delta\Psi$), whereas the gravitational potential contribution is free from $\partial_i^n\Psi$ for any $n$. These contributions to the luminosity distance are explicitly 
\begin{align}
\label{Eq:components}
	\delta \mathcal D_L^V =&~\left(1-\frac{1}{\mathcal H_z\bar r_z}\right)V_{\|s} +\frac{1}{\mathcal H_z\bar r_z}V_{\|o}\, ,\nonumber\\
	\delta \mathcal D_L^\text{lens.} =&-\int_0^{\bar r_z}d\bar r \left\{(\bar r_z-\bar r)\frac{\bar r}{\bar r_z}\Delta\Psi \right\}
	=-\frac{3}{2}\mathcal H_o^2\Omega_{m}\int_0^{\bar r_z}d\bar r \left\{(\bar r_z-\bar r)\frac{\bar r}{\bar r_z}\frac{1}{a(\bar\tau_o-\bar r)} \delta( \bar\tau_o-\bar r, \bar r\hat n)\right\}	 \, ,\nonumber\\
	\delta \mathcal D_L^{\Psi} =&~ \left(\mathcal H_o+\frac{1}{\bar r_z}-\frac{\mathcal H_o}{\mathcal H_z\bar r_z}\right)\delta\tau_o +\frac{1}{\mathcal H_z\bar r_z}\Psi_s-\frac{1}{\mathcal H_z\bar r_z}\Psi_o-\Psi_s\nonumber\\
	&+\int_0^{\bar r_z}d\bar r \left\{ \frac{2}{\bar r_z}\Psi +2\left( \frac{1}{\mathcal H_z\bar r_z}+\frac{\bar r}{\bar r_z}-1\right)\Psi'+\left(\bar r_z-\bar r\right)\frac{\bar r}{\bar r_z}\Psi''	\right\}\, ,
\end{align}
where the suffix $o$ and $s$ indicate that a variable is evaluated at the observation and the source, respectively. Note that we used the Poisson equation $\Delta \Psi  =\frac{3}{2}\mathcal H_o^2\Omega_{m} \delta/a$ in $\delta\mathcal D_L^\text{lens.}$, where $\delta$ is the linear-order matter density contrast in the Newtonian cosmology  or the synchronous gauge. One can decompose the velocity and the potential into time-dependent and scale-dependent parts as $V_\|(\tau,\bm x) = D_V(\tau)\partial_\|\zeta(\bm x)$, $\Psi(\tau,\bm x) = D_\Psi(\tau)\zeta(\bm x)$, where $\zeta$ is the time-independent but scale dependent curvature perturbation. The  solutions of $D_V$ and $D_\Psi$ in Eq. (\ref{Eq:components}) are presented in~\cite{BIYO16,YOGO16}. The detailed derivations of these equations are  presented in~\cite{SASAK87,YOO14a,SCJE12}.

\section{Analytic expressions for the two-point correlation function of the luminosity distances}
\label{Sec:correlation}

\begin{figure}[t]
\centering
\includegraphics[width=0.8\textwidth]{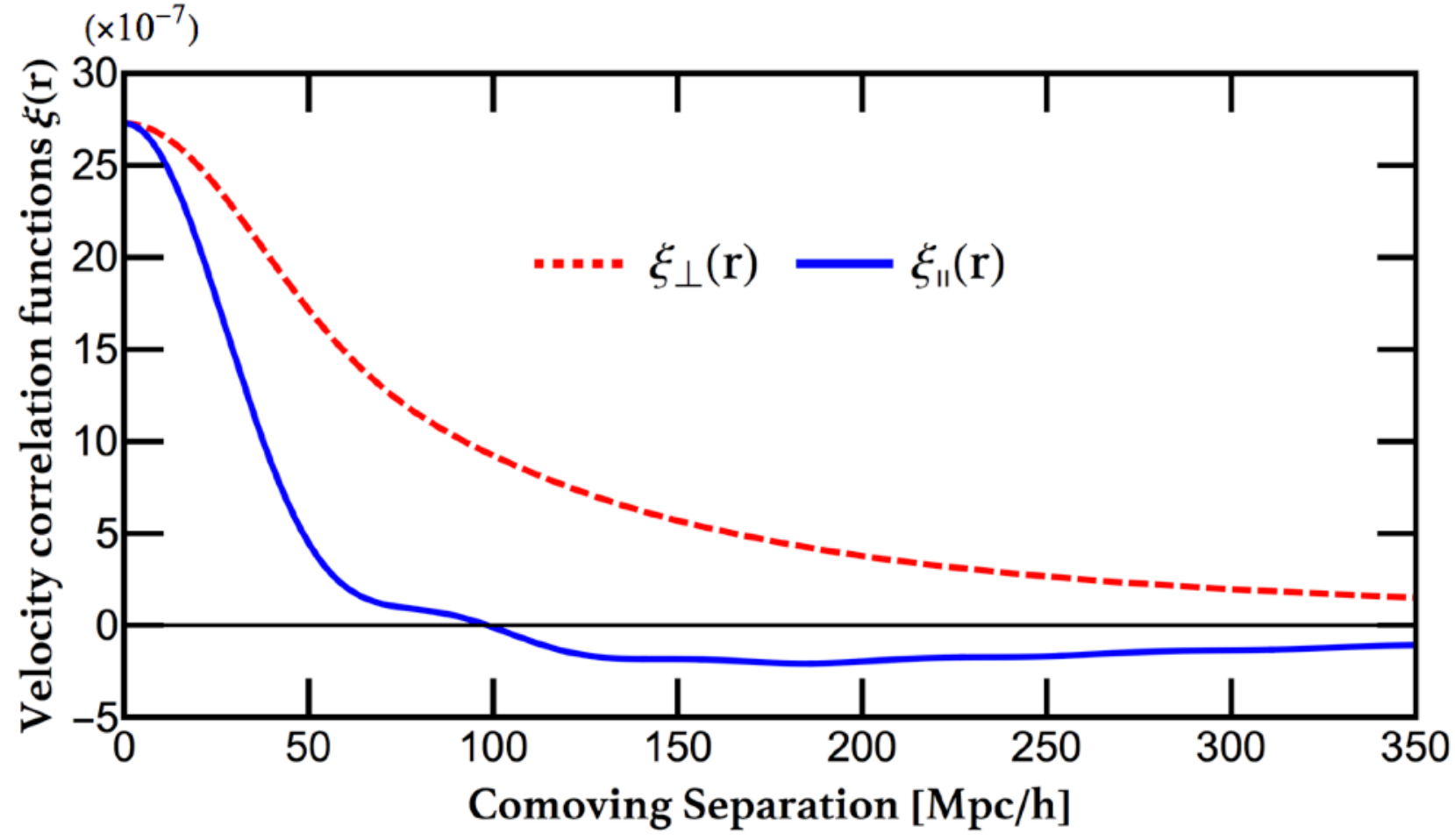}
\caption{The velocity correlation function $\left<V_\|(z_1,\hat n_1)V_\|(z_2,\hat n_2)\right>$ is decomposed as the correlations $\xi_\|$ and $\xi_\perp$ parallel and perpendicular to the separation vector, respectively (see Eq. (\ref{Eq:vel_comp})). The computations are performed under the choices of $k_\text{IR}=\mathcal H_o$ and $k_\text{UV}=0.1~h/\text{Mpc}$. }
\label{Fig:vel_corr}
\end{figure}

%
%
To derive the velocity contribution to the correlation function of the luminosity distances, we need the two-point correlation function of radial velocities: $\left<V_\|(z_1,\hat n_1)V_\|(z_2,\hat n_2)\right>$. Using $V_\| (\tau,\bm x)= D_V(\tau) \partial_\|\zeta(\bm x)$, the correlation function of the radial velocities can be expressed as
\begin{align}
\label{Eq:vel_corr}
\left<V_\|(z_1,\hat n_1)V_\|(z_2,\hat n_2)\right>=\hat n^i_1\hat n^j_2  \left<V_i(z_1,\hat n_1)V_j(z_2,\hat n_2)\right> 	=\left(\frac{\mathcal C}{\mathcal H_o}\right)^2 D_{V_1}D_{V_2}\left\{\hat{\mathcal P}_\|\xi_{\|}(|\bm r|)+\hat{\mathcal P}_\perp \xi_{\perp}(|\bm r|)\right\}\, ,
\end{align}
where the suffix $i$ indicates that a temporal function is evaluated at $z_i$ e.g., $D_{V_1}=D_V(z_1)$, and $\mathcal C\equiv -\frac{5}{2}\mathcal H_o^2\Omega_m$.  Note that we decompose the velocity correlation function $\left<V_iV_j\right>$ into the parallel and the perpendicular components with respect to the separation vector $\bm r=\bar r_{z_1}\hat {\bm{n}}_1-\bar r_{z_2}\hat {\bm{n}}_2$: $\left<V_iV_j\right> = \hat r_i\hat r_j\xi_{\|}+\left(\delta_{ij}-\hat r_i\hat r_j\right)\xi_{\perp}$, and we defined $\hat{\mathcal P}_\| \equiv \hat n_1^i\hat n_2^j \hat r_i\hat r_j$, and $\hat{\mathcal P}_\perp \equiv \hat n_1^i\hat n_2^j (\delta_{ij}-\hat r_i\hat r_j)$.  The correlation functions of the perpendicular and the parallel components $\xi_\perp$ and $\xi_{\|}$ are defined respectively as
\begin{equation}
\label{Eq:vel_comp}
	\xi_\perp(r)\equiv- \mathcal H_o^2\int_{k_\text{IR}}^{k_\text{UV}} \frac{dk}{2\pi^2}P_m(k)\frac{j_0'(kr)}{kr}\, ,~~~~~~~~~~~~~~\xi_\|(r)\equiv- \mathcal H_o^2\int_{k_\text{IR}}^{k_\text{UV}} \frac{dk}{2\pi^2}P_m(k){j_0''(kr)}\, ,
\end{equation}
where $j_0(x)$ is the spherical Bessel function, and $P_m(k)$ is the  matter power spectrum   in the synchronous gauge  computed by using $\texttt {CAMB}$.  $k_\text{IR}$ and $k_\text{UV}$ represent the lower and the upper cutoffs in the integration, respectively. Figure~\ref{Fig:vel_corr} shows $\xi_\perp$ and $\xi_\|$ with respect to the separation $r$. Note that $\hat{\mathcal P}_\perp$ vanishes when the line-of-sights of two sources are aligned, and the contribution of $\hat{\mathcal P}_\perp$ is  the characteristic of the case $\hat {\bm n}_1\neq \hat {\bm n}_2$ i.e., the wide angle effect.

Utilizing Eq. (\ref{Eq:vel_corr}), one can derive the auto-correlation function of the velocity contribution: 
\begin{align}
\label{Eq:velocity}
\left<\delta\mathcal D_L^V(z_1,\bm{\hat n_1}) \delta\mathcal D_L^V(z_2,\bm{\hat n_2})\right>
&=~\left(\frac{\mathcal C}{\mathcal H_o}\right)^2D_{V_1}D_{V_2}\left(1-\frac{1}{\mathcal H_{z_1}\bar r_{z_1}}\right) \left(1-\frac{1}{\mathcal H_{z_2}\bar r_{z_2}}\right) \Big[ \hat {\mathcal P}_\perp \xi_{\perp}(|\bm r|)+\hat {\mathcal P}_\|\xi_\|(|\bm r|) \Big]\nonumber\\
&~+\left(\frac{\mathcal C}{\mathcal H_o}\right)^2D_{V_1}D_{V_o}\frac{1}{\mathcal H_{z_2}\bar r_{z_2}}\left(1-\frac{1}{\mathcal H_{z_1}\bar r_{z_1}}\right) \hat{\mathcal P}_\|  \xi_\|(\bar r_{z_1})+\left(1\leftrightarrow2\right) \nonumber\\
&~+\mathcal C^2 D_{V_o}^2\frac{1}{(\mathcal H_{z_1}\bar r_{z_1})}\frac{1}{(\mathcal H_{z_2}\bar r_{z_2})} \hat{\bm n}_1\cdot \hat{\bm n}_2  \int_{k_\text{IR}}^{k_\text{UV}}\frac{dk}{6\pi^2}P_m(k)\, .
\end{align}
%


 \noindent
As shown in Fig. \ref{Fig:vel_corr}, the velocity correlation $\xi_\perp$ perpendicular to the separation is always positive, and it decays monotonically as the separation increases. The velocity correlation $\xi_\|$ parallel to the separation also decreases monotonically as the separation increases before reaching to the sign flip ($r\simeq 100~\text{Mpc}/h$) from positive to negative, and it is nearly independent of the separation beyond the sign flip.   Note that $\xi_\|$ and $\xi_\perp$ are derived from the cutoff choices $k_\text{IR}=\mathcal H_o$ and $k_\text{UV}=0.1~h/\text{Mpc}$. The super-horizon modes contribution to $\xi_\|$ and $\xi_\perp$ is negligible, and we choose $k_\text{IR}=\mathcal H_o$. In contrast, for larger choice of $k_\text{UV}$ e.g., $k_\text{UV}= 1~h/\text{Mpc}$, the wiggle of $\xi_\|$ in Fig. \ref{Fig:vel_corr} disappears, and $\xi_\|$ becomes a smooth function, while the shape of $\xi_\perp$ does not change since the integrand of $\xi_\perp$ in Eq. (\ref{Eq:vel_comp}) is insensitive on large $k$. In addition to the shape-changing, in the case of the cutoff choice, the maximum values of $\xi_\|$ and $\xi_\perp$ are a factor of 1.5 larger than those in Fig. \ref{Fig:vel_corr}. However, the larger choice $k_\text{UV}>1~h/\text{Mpc}$ does not yield substantial numerical difference to $\xi_\|$ and $\xi_\perp$.

 The first line in Eq. (\ref{Eq:velocity}) is the auto-correlation of the source velocity contribution. Because of the property of $\xi_\|$ that becomes negative for large separations, the source velocity contribution to the correlation function of the luminosity distances can be anti-correlated (the directions of each velocity are opposite) when the separation is large enough and $\hat{\mathcal P}_\perp \simeq 0$. However, in other cases, the sign of this correlation function is positive. 

 The second line corresponds to the cross-correlation of the contributions of the velocities at source and observer positions. It vanishes when two sources are far away from the observer but with small angular separation i.e., $\hat{\mathcal P}_\|\simeq 0$. ($\hat n_1  \mathbin{\!/\mkern-5mu/\!}\hat n_2\perp \hat r$). However, since the comoving separation is large $\bar r_{z_1},\bar r_{z_2}>300~\text{Mpc}/h$ for $z_1,z_2>0.1$, there is in practice a strong suppression in the integrand due to the suppression the spherical Bessel function $j_0(k\bar r_z)$, regardless of $\hat{\mathcal P}_\|$. Thus, the magnitude of the cross-correlation is much smaller than the correlation functions of the other velocity contributions.
 
  The last line represents the auto-correlation of the contribution of the velocity at observer position, and it depends on the angular separation between sources and the source distances from the observer. Since this line is proportional to the correlation of the observer velocity components along two light-of-sights of sources, it vanishes when the angular separation between sources is $90~{}^\circ$ i.e., $\hat {\bm n}_1\cdot \hat {\bm n}_2=0$ and increases as the angular separation decreases. In addition, it has a small value when the sources are far away from the observer. Apparent in Eq.~(\ref{Eq:components}), the observed luminosity distance depends  on the velocity contribution at the observer position, and its ensemble average is used for computing the correlation in Eq.~(\ref{Eq:velocity}). The ergodic theorem says our observational average is equivalent to the average over many realizations of the Universe, and Eq.~(\ref{Eq:velocity}) represents we make observations over many realizations of the Universe, in which the velocity at the observer position changes.

The other important contribution is the gravitational lensing contribution. The lensing correlation function is given by 
\begin{align}
\label{Eq:lensing}
\left<\delta\mathcal D_L^\text{lens.}(z_1,\bm{\hat n_1})\delta\mathcal D_L^\text{lens.}(z_2,\bm{\hat n_2})\right>
=&~\frac{9}{4}\mathcal H_o^4\Omega_m^2\int_0^{\bar r_{z_1}}d\bar r_1 g(\bar r_1)\int_0^{\bar r_{z_2}}d\bar r_2 g(\bar r_2) ~\xi_m\Big(\left|\bar r_1 \hat{\bm n}_1-\bar r_2 \hat{\bm n}_2 \right|\Big)\, ,
\end{align}
where $g(\bar r_i)\equiv \frac{(\bar r_{z_i}-\bar r_i)\bar r_i}{\bar r_{z_i}} \frac{D(\bar\tau_o-\bar r_i)}{a(\bar\tau_o-\bar r_i)}$, and $\xi_m(|\bm r|)$ is the matter correlation function i.e., $\xi_m(|\bm r|) =\left<\delta(\bm x)\delta(\bm x+\bm r)\right>$, which decays as the separation $|\bm r|$ increases. Since we shall consider  large angular separation cases, we will compute this contribution accurately instead of applying the Limber approximation, which works well only for small angular separation~(e.g., see \cite{SI06}).


Our numerical integration of the lensing correlation is performed along the paths of two photons emitted from two sources, and it normally increases as the distance between the observer and the source increases. However, for a large angular separation, the gravitational lensing contribution becomes quickly negligible.  The reason is that the separation of the matter correlation function  increases quickly with respect to $\bar r_1$ and $\bar r_2$ when the angular separation is large. That is, the matter correlation function in the integration decays sharply as $\bar r_1$ and $\bar r_2$ increase when the angular separation is large. Thus, the lensing correlation can be mostly significant for small angular separation and large redshift.  
\begin{figure}[h!]
\centering
\includegraphics[width=0.8\textwidth]{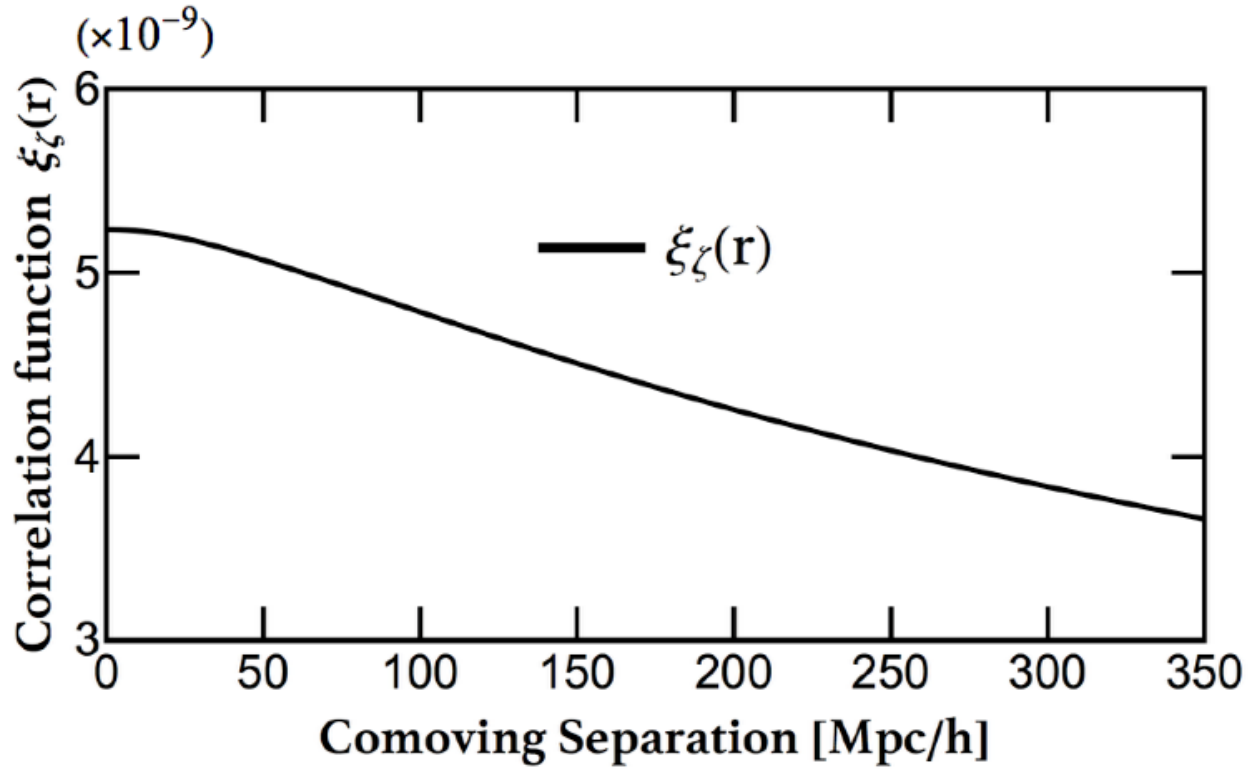}
\caption{The auto- correlation function $\xi_\zeta(r)$ of the curvature perturbation in the $\Lambda$CDM universe ($\Omega_m=0.3$) for $k_\text{IR}=\mathcal H_o$ and $k_\text{UV}=0.1~\text{Mpc}/h$ (see appendix~\ref{App:GP} for details). }
\label{Fig:zeta_corr}
\end{figure}
%


Finally, we discuss the gravitational potential contribution to the luminosity distance correlation.\footnote{Since the expression of the auto-correlation of the gravitational potential contribution is complicated, we present the explicit expression in appendix~\ref{App:GP}.}   The auto-correlation function of the gravitational potential is $\left<\Psi(z_1,\hat n_1)\Psi(z_2,\hat n_2)\right>=D_\Psi(z_1)D_\Psi(z_2)\xi_\zeta(|\bm r|)$,  and the correlation function $\xi_\zeta$ of the curvature perturbation in Fig.~\ref{Fig:zeta_corr} is
\begin{equation}
	\xi_\zeta(|\bm r|)\equiv \left<\zeta(\bm x)\zeta(\bm x+\bm r)\right>=\frac{25}{4}\mathcal H_o^4\Omega_m^2\int_{k_\text{IR}}^{k_\text{UV}} \frac{dk}{2\pi^2}\frac{1}{k^2}P_m(k)j_0(kr)\, .
\end{equation}
Given the red tilt of the spectral index, the correlation function (and its variance) $\xi_\zeta$ diverges as we include modes on larger scales ($k<k_\text{IR}$). However, as shown in~\cite{BIYO16}, the contributions of the gravitational potential at $k<k_\text{IR}$ cancel in $\delta\mathcal D_L$, and their exact value is rather insensitive as long as $k_\text{IR}<\mathcal H_o$. Figure~\ref{Fig:zeta_corr} and our numerical calculations of the gravitational potential represent their effective contributions to $\delta\mathcal D_L$ with the long-mode cancellation taken into account.

 As depicted in Fig.~\ref{Fig:zeta_corr}, the auto-correlation function of the curvature perturbation decreases as the separation increases, but the amplitude does not change significantly. Thus, the auto-correlation function of the gravitational potential contribution is nearly independent of separation, and its amplitude is similar to the amplitude of the luminosity distance variance of this contribution.  Since $D_\Psi$ is nearly time-independent in a $\Lambda$CDM universe, by assuming $D_\Psi(z)=D_{\Psi o}$ we can simplify the gravitational potential contribution for sufficiently large redshift as $\delta\mathcal D_L^\Psi \simeq \left(1-\frac{1}{\mathcal H_z\bar r_z}\right)\mathcal H_o\delta \tau_o+2 \Psi_o = \left(1-\frac{1}{\mathcal H_z\bar r_z}\right)\left(\Psi_o+\zeta_o\right)+2 \Psi_o$. From this, we can derived the auto-correlation function of the simplified one: 
\begin{equation}
\label{Eq:GP_app}
	\left<\delta\mathcal D_L^\Psi (z_1,\hat n_1)\delta\mathcal D_L^\Psi (z_2,\hat n_2)\right> \simeq \left\{\left(2D_{\Psi o}+1 -\frac{1}{\mathcal H_{z_1}\bar r_{z_1}}\left(D_{\Psi o}+1\right)\right)\times \left(1\leftrightarrow2\right)\right\}\xi_\zeta(0)\, .
\end{equation}
The numerical value of the redshift dependent part in Eq.~(\ref{Eq:GP_app}) is $3.7$ ($0.03$) at $z_1=z_2=0.5$ ($z_1=z_2=1.0$). That is, the amplitude of the auto-correlation function of the gravitational potential contribution is approximately $\mathcal O\left(10^{-8}\right)$ ($\mathcal O\left(10^{-10}\right)$) at $z_1=z_2=0.5$ ($z_1=z_2=1.0$). We shall see the detailed numerical results in the following section.

%
%

\section{Results}
\label{Sec:results}
In this section we present the numerical computations of the velocity, the gravitational potential, and the lensing contributions to the correlation function of the luminosity distances. Especially, we discuss how the angular correlation function of the luminosity distances depends on the matter content today in a flat $\Lambda$CDM universe.

In principle, the integrations in the expression of the luminosity distance correlation function should be performed from zero-wavenumber to infinite-wavenumber i.e., $k_\text{IR}=0$ and $k_\text{UV}=\infty$. However, as shown in \cite{BIYO16}, the contribution of super-horizon perturbation modes to the luminosity distance is negligible, if we use the gauge-invariant expression of the luminosity distance.   Therefore, there is no significant numerical difference between the choices $k_\text{IR}=\mathcal H_o$ and $k_\text{IR}<\mathcal H_o$, and we set the lower cutoff as $k_\text{IR}=\mathcal H_o$. In the case of the upper cutoff choice, we set mostly $k_\text{UV}=0.1~h/\text{Mpc}$ since the nonlinearity becomes important at $k\gtrsim 0.1~h/\text{Mpc}$. In fact, the larger $k_\text{UV}$ yields the larger amplitude at a certain range $0.1~h/\text{Mpc}\lesssim k_\text{UV}\lesssim 1~h/\text{Mpc}$, and the upper cutoff choice does not yield significant difference at $k_\text{UV} \gtrsim 1~h/\text{Mpc}$ (see~\cite{BEGAET12a}). We also discuss how numeric amplitudes depend on a choice of $k_\text{UV}$.




 \subsection{Angular correlation function of the luminosity distance ($z_1=z_2$)}
 \label{Sec:angular_correlation}

\begin{figure}[t]
\centering
\includegraphics[width=1\textwidth,height=0.4\textwidth]{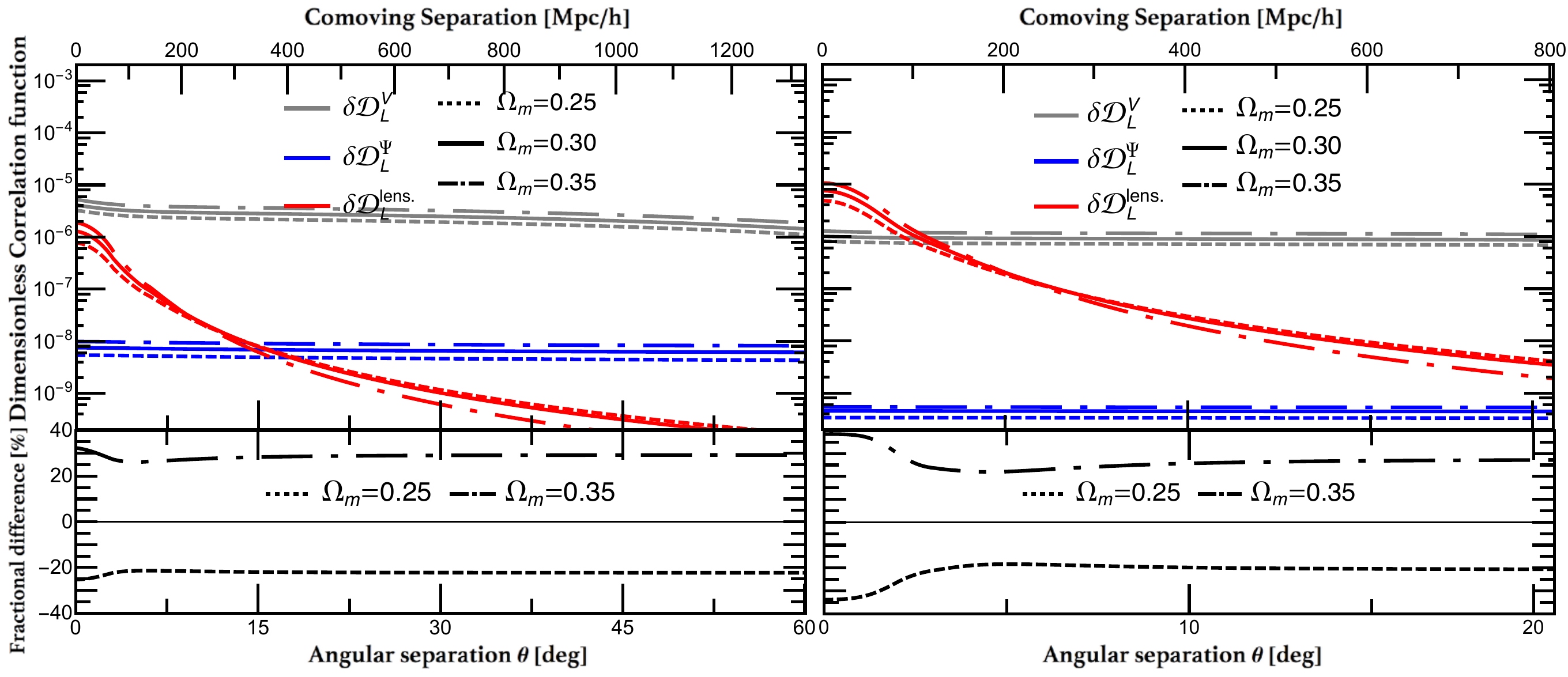}
\caption{ The left-top (right) figure shows the (dimensionless) angular correlation functions of $z=0.5$ ($z=1.0$), and the corresponding fractional differences between $\Omega_m=0.3$ and others ($\Omega_m=0.25~\text{and}~0.35$) are illustrated in the bottom figures.The grey, the blue, and the red lines indicate the velocity, the gravitational potential, and the lensing contributions, respectively. Different line types represent the different cosmological parameter sets: the dotted, the solid, and the dot-dashed lines corresponds to $\Omega_m=0.25$, $\Omega_m=0.3$, and $\Omega_m=0.35$ in the flat $\Lambda$CDM universe, respectively. The conversion of the angle $\theta$ to the comoving separation is made by assuming $\Omega_m=0.3$.}
\label{Fig:angular}
\end{figure}
%
 
 The angular correlation functions of the three contributions at $z=0.5$ ($z=1.0$) are illustrated in the upper left (right) figure in Fig. \ref{Fig:angular}. The bottom panels show the  fractional difference between the $\Omega_m=0.3$ universe and others ($\Omega_m=0.25$ and $\Omega_m=0.35$). Note that the range of the angular separation is different in the left and the right panels since the coverage of a deep survey at high redshift is usually narrow.
  

The velocity contribution to the angular correlation function is important at redshift $z=0.5$ for any angular separation, and at redshift $z=1.0$ for $\theta \gtrsim 5^\circ$.   The lensing  contribution is dominant at $z=1.0$ but only for small angular separation $\theta\lesssim 5^\circ$ since it decays sharply as the angular separation increases compared to the other contributions.  As discussed in section~\ref{Sec:correlation}, the gravitational potential contribution to the luminosity distance correlation function is nearly independent of the  separation between sources, and its amplitudes are approximately $\mathcal O\left(10^{-8}\right)$ at $z_1=z_2=0.5$ and $\mathcal O\left(10^{-10}\right)$ at $z_1=z_2=1.0$. As a result, the gravitational potential contribution is always smaller than the other contributions, except for the lensing contribution at large separation. As shown in Fig. \ref{Fig:angular}, the larger $\Omega_m$ yields the larger amplitude for the velocity contribution over the angular scales. However, the effect of $\Omega_m$ on the lensing contribution is more complicated. Similarly to the velocity contribution, the lensing correlation function also increases as $\Omega_m$ increases but only for small angular separation, and this tendency becomes opposite when the angular separation is large. In the velocity contribution to the luminosity distance correlation function, the dominant contribution is the auto-correlation of the velocity at the observer's position (the third line in Eq. (\ref{Eq:velocity})), and it is proportional to $(\mathcal CD_{Vo})^2$, where $\mathcal C D_V = -\frac{d\ln D}{d\ln a} \mathcal H D$. In a $\Lambda$CDM universe, $\frac{d\ln D}{d\ln a}$ is simply expressed as $\frac{d\ln D}{d\ln a}\simeq\Omega_m^{0.55}$, and the $(\mathcal C D_{Vo})^2$ is proportional to $\Omega_m^{1.1}$. While the velocity contribution is proportional to $\Omega_m^{1.1}$, the lensing correlation is proportional to $\Omega_m^2$ as presented in Eq. (\ref{Eq:lensing}). As a result, the fractional difference between the universe with $\Omega_m=0.3$ and the other cases ($\Omega_m=0.25$ and $\Omega_m=0.35$) is enhanced when the lensing contribution is significant (small angular separation and high redshift). 

  When we choose larger $k_\text{UV}$ e.g., $k_\text{UV}=10~h/\text{Mpc}$, the amplitude of the auto-correlation function of the contribution of the velocity at the observer increases with a factor of 1.5 as discussed previously in section~\ref{Sec:correlation}, and this contribution is independent from the separations of sources. Thus, as $k_\text{UV}$ increases from $k_\text{UV}=0.1~h/\text{Mpc}$ to $k_\text{UV}=1~h/\text{Mpc}$, the amplitude of $\left<\delta D_L^V \delta D_L^V\right>$ increases at all separations until the angular separation reaches $90^\circ$. In addition, the lensing contribution is more sensitive on a larger $k_\text{UV}$ choice, for instance, the maximum amplitude of $\left<\delta D_L^\text{lens.} \delta D_L^\text{lens.} \right>$ with $k_\text{UV}=1~h/\text{Mpc}$ ($k_\text{UV}=10~h/\text{Mpc}$) is about 3.5 (5) times larger than that with $k_\text{UV}=0.1~h/\text{Mpc}$ at $z=0.5$ and $z=1.0$. However, in contrast to the velocity contribution case, the amplitude of the lensing correlation increases just at small separations ($\theta \lesssim 3^\circ$) with respect to $k_\text{UV}$. At large separations, the change is insignificant, and the amplitude of the lensing contribution is also subdominant to the velocity contribution at any redshift.




 \subsection{Three-dimensional two-point correlation function of the luminosity distance ($z_1\neq z_2$)}
\label{Sec:2pt_correlation}

\begin{figure}[h]
\centering
\includegraphics[height=0.4\textwidth,width=0.4\textwidth]{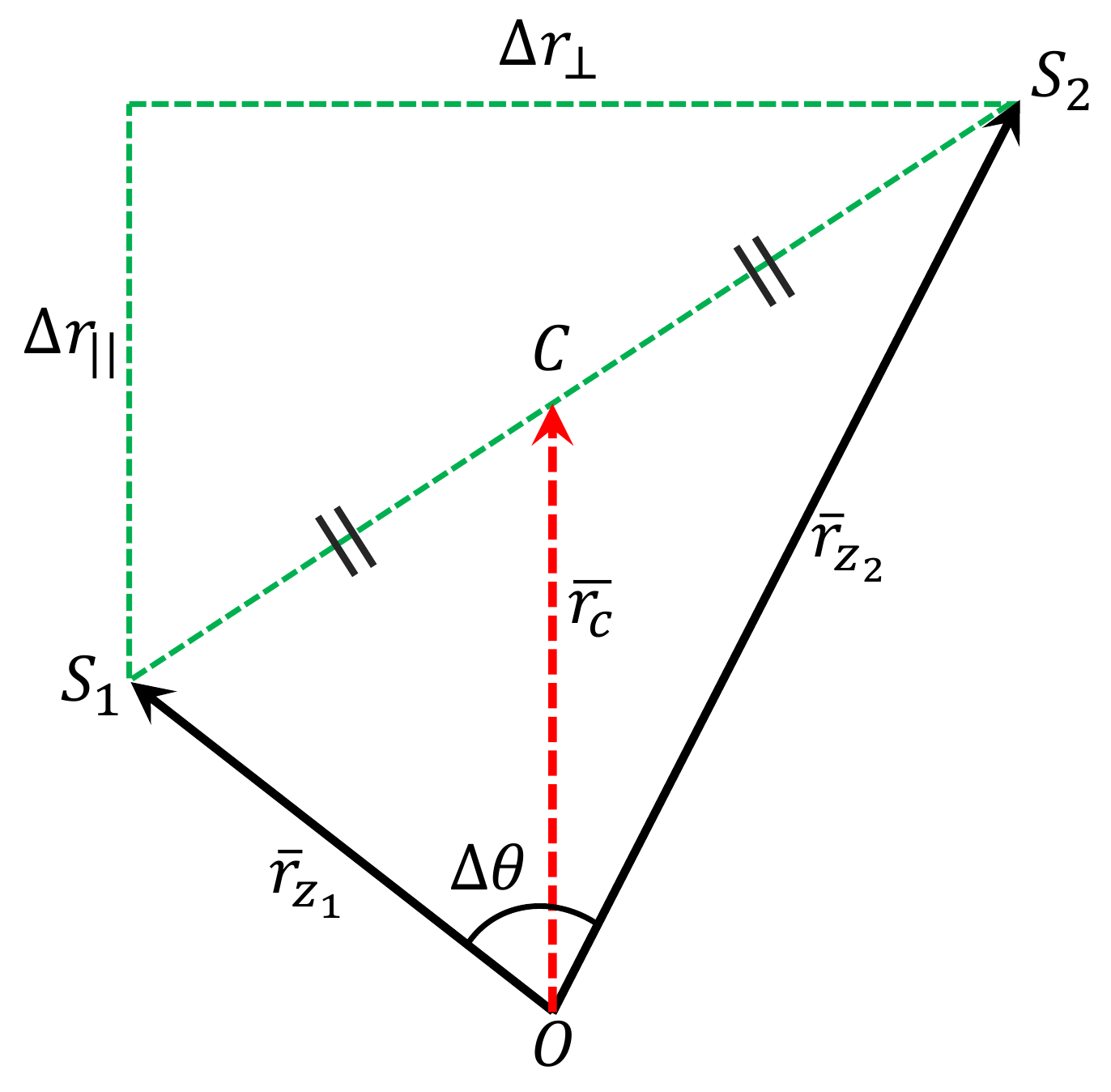}
\caption{  Points $S_1$ and $S_2$ denote the source positions whose vectors from the observation position $O$ are $\bar{\bm r}_{z_1}$ and $\bar{\bm r}_{z_2}$, respectively. $\Delta\theta$ is the angle between these two vectors. The point $C$ is the midpoint which equally divides the connecting line between $S_1$ and $S_2$, and the midpoint vector is $\bar{\bm r}_c$. $\Delta r_\|$ ($\Delta r_\perp$) indicates the parallel (perpendicular) separation with respect to the midpoint vector $\bar{\bm r}_c$.}
\label{Fig:midpoint}
\end{figure}

As discussed, the correlation functions of the velocity and the lensing contribution are anisotropic. For this reason, we present the correlation function as a function of the parallel and the perpendicular separations. These separations are defined in the midpoint coordinate illustrated in Fig. \ref{Fig:midpoint} (see also~\cite{YNKBN06,SABRPE15,BGRP15}). As depicted in Fig. \ref{Fig:midpoint}, the midpoint vector $\bm{ r_c}$ points to the midpoint on the separation of the sources i.e., $\bm{ r_c} = \frac{1}{2}\left(\bar r_{z_1}\hat{ \bm n}_1+\bar r_{z_2} \hat {\bm n}_2\right)$, where $\bar r_{z_1}$ and $\bar r_{z_2}$ are the comoving distances of the sources. Two distance vectors can be expressed in terms of the parallel separation $\Delta r_\|$, the perpendicular separation $\Delta r_\perp$, and the midpoint distance from the observer $\bar r_c$ by utilizing the Heron's formula and the geometric length of $\overline{OS_1}$ as\footnote{
The Heron's formula is 
\begin{equation}
	A = \sqrt{s(s-a)(s-b)(s-c)}\, ,
\end{equation}
where $A$ is the area, $a$, $b$, and $c$ are lengths of each side. $s$ is defined as $s\equiv  \frac{a+b+c}{2}$.} 
\begin{eqnarray}
	\label{Eq:r1r2}
	~~\bar r_{z_1} = \frac{1}{2}\sqrt{(2\bar r_c-\Delta r_\|)^2+\Delta r_\perp^2},~~~~~~~~~~~~~~\bar r_{z_2} &=& \frac{1}{2}\sqrt{(2\bar r_c+\Delta r_\|)^2+\Delta r_\perp^2} 	\, ,
\end{eqnarray}
\begin{figure}[t]
\includegraphics[height=0.5\textwidth,width=1\textwidth]{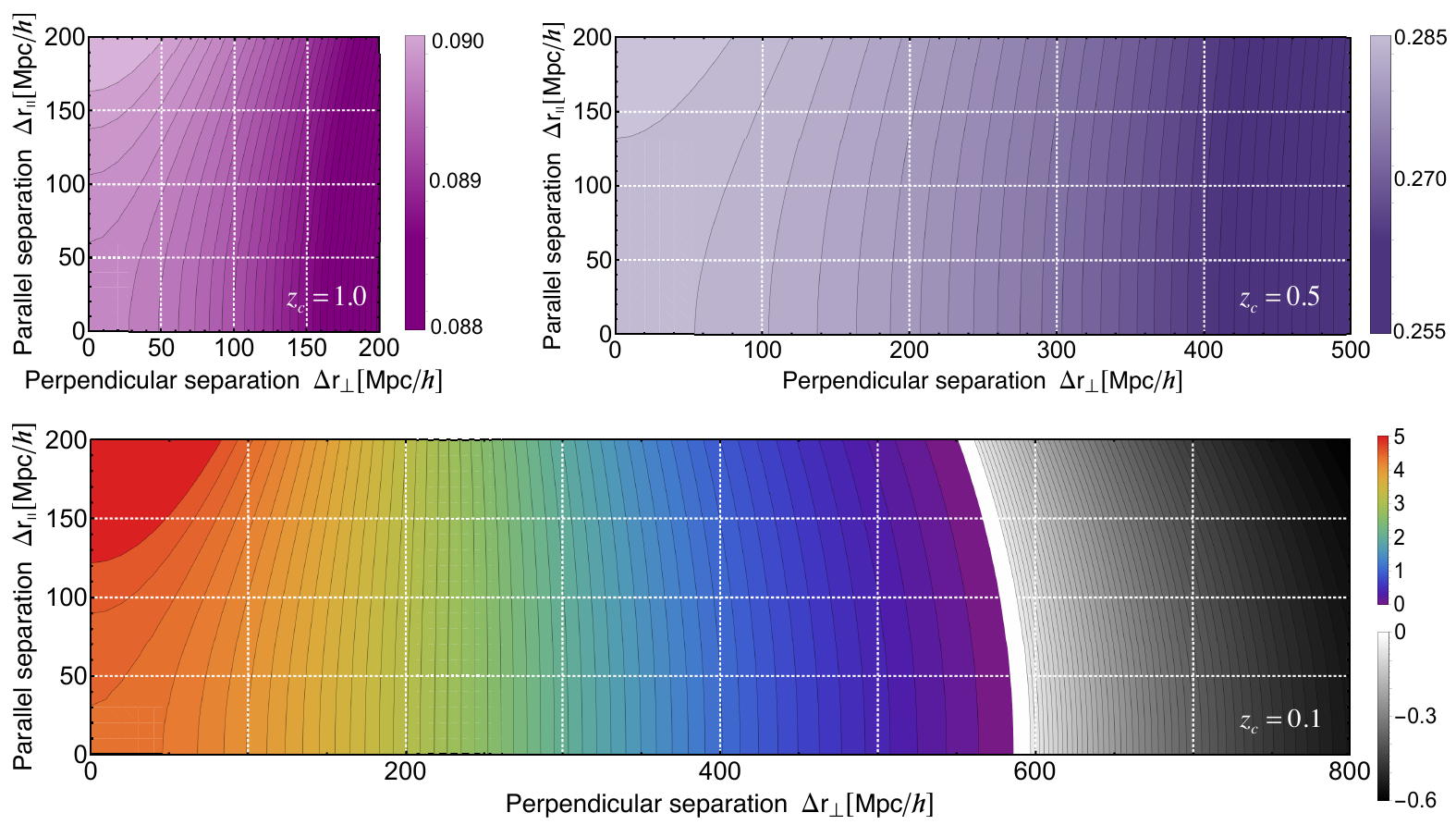}
\caption{Auto-correlation of the observer velocity contribution of the flat $\Lambda$CDM universe of $\Omega_m=0.3$. The amplitudes are multiplied by $10^5$ e.g., $5$ in the figure means $5\times 10^{-5}$. The rainbow (gray) color indicates the positive (negative) correlation. The navy ($\bar z_c=0.5$) and the purple ($\bar z_c=1.0$) unicolors with gradations describe subtle changes of their amplitude with respect to $\Delta r_\|$ and $\Delta r_\perp$. The result with the midpoint redshift $\bar z_c=0.1$ is shown in the bottom, and that with $\bar z_c=0.5$ is presented in the right upper figure. The left top figure corresponds to that with $\bar z_c=1.0$. }
\label{Fig:vel_obs}
\end{figure}
 The angular separation $\Delta\theta$ is determined from the geometric length between two sources  as
\begin{equation}
\label{Eq:angle}
	\cos\Delta\theta = \frac{4\bar r_c^2-\Delta r_\|^2-\Delta r_\perp^2}{\bar r_{z_1}\bar r_{z_2}}		\, .
\end{equation}
For instance, $\hat{\mathcal P}_\perp$ and $\hat{\mathcal P}_\|$ in Eq. (\ref{Eq:velocity}) are expressed in terms of $\bar r_c$, $\Delta r_\|$, and $\Delta r_\perp$ as
\begin{align}
	\hat{\mathcal P}_\|(\bar r_z,\Delta r_\|,\Delta r_\perp) =&~( \hat n_1\cdot\hat r)(\hat n_2\cdot\hat r) = \frac{\bar r_{z_1}-\bar r_{z_2}\cos\Delta\theta}{\sqrt{\bar r_{z_1}^2+\bar r_{z_2}^2-2\bar r_{z_1}\bar r_{z_2}\cos\Delta\theta}}\times \frac{\bar r_{z_1}\cos\Delta\theta-\bar r_{z_2}}{\sqrt{\bar r_{z_1}^2+\bar r_{z_2}^2-2\bar r_{z_1}\bar r_{z_2}\cos\Delta\theta}}\nonumber\\	
	  =&-\frac{\left(\Delta r_\|^2+\Delta r_\perp^2-2\bar r_c\Delta r_\|\right)\left(\Delta r_\|^2+\Delta r_\perp^2+2\bar r_c\Delta r_\|\right)}{\left(\Delta r_\|^2+\Delta r_\perp^2\right) \sqrt{ \Delta r_\perp^2+(\Delta r_\|-2\bar r_c)^2}\sqrt{ \Delta r_\perp^2+(\Delta r_\|+2\bar r_c)^2} } \, ,\nonumber\\	
	\hat{\mathcal P}_\perp(\bar r_z,\Delta r_\|,\Delta r_\perp) =&~\hat n_1\cdot\hat n_2-\hat{\mathcal P}_\| = \cos\Delta\theta-\hat{\mathcal P}_\|\nonumber\\
	 =&~ \frac{4\bar r_c^2 \Delta r_\perp^2}{\left(\Delta r_\|^2+\Delta r_\perp^2\right) \sqrt{ \Delta r_\perp^2+(\Delta r_\|-2\bar r_c)^2}\sqrt{ \Delta r_\perp^2+(\Delta r_\|+2\bar r_c)^2} } \, .
	 	\end{align}
%

We numerically investigate the three-dimensional two-point correlation function of the luminosity distances for given midpoint distances: $\bar r_c\simeq 293~\text{Mpc}/h$, $1323~\text{Mpc}/h$, and $2314~\text{Mpc}/h$, corresponding to redshifts $ \bar z_c=0.1$, $\bar z_c=0.5$, and $ \bar z_c =1.0$ in a flat $\Lambda$CDM universe with $\Omega_m=0.3$. The wide angle effect, which originates from the difference in the two line-of-sights of sources, is taken into account in the results.  The predicted noise $\sigma_\text{mm}$ on the correlation  is presented in~\cite{SNMB16} as
\begin{equation}
	\sigma_\text{mm} = \frac{\sigma_\text{err}^2}{\sqrt{N_p}}\, , 
\end{equation}
where $\sigma_\text{err}$ is the overall uncertainty on individual supernova magnitude measurements, and $N_p$ is the number of pairs within a certain angular separation $[\theta,\theta+\delta\theta]$. The explicit expression of $N_p$ is $N_p=N_s(N_s-1)\pi \theta\Delta\theta/A_\text{survey}$, where $N_s$ is the number of source which is uniformly distributed on the survey area $A_\text{survey}$. With $\sigma_\text{mm}$, the signal-to-noise (SNR) of the correlation function of the luminosity distances becomes
\begin{equation}
	\text{SNR}=\frac{\left<\delta\mathcal D_L(z_1,\hat n_1)\delta\mathcal D_L(z_2,\hat n_2)\right>}{\sigma_\text{mm}}\, .
\end{equation}
In the case of the deep LSST survey, the number of source and the survey area are $N_s\simeq 10^4$, $A_\text{survey}=20~\text{deg}^2$, and the overall uncertainty in current observation is $\sigma_\text{err}\simeq 0.1$. Thus, the predicted noise in this case is  $\sigma_\text{mm}\sim 10^{-5}$ when the angular separation and the angular bin size are approximately $\theta=5^\circ$ and $\delta\theta=0.01^\circ$, respectively. To compare the amplitudes of the correlation function and of the estimated noise, the amplitudes of figures are multiplied by $10^5$.

\begin{figure}[t]
\includegraphics[height=0.5\textwidth,width=1\textwidth]{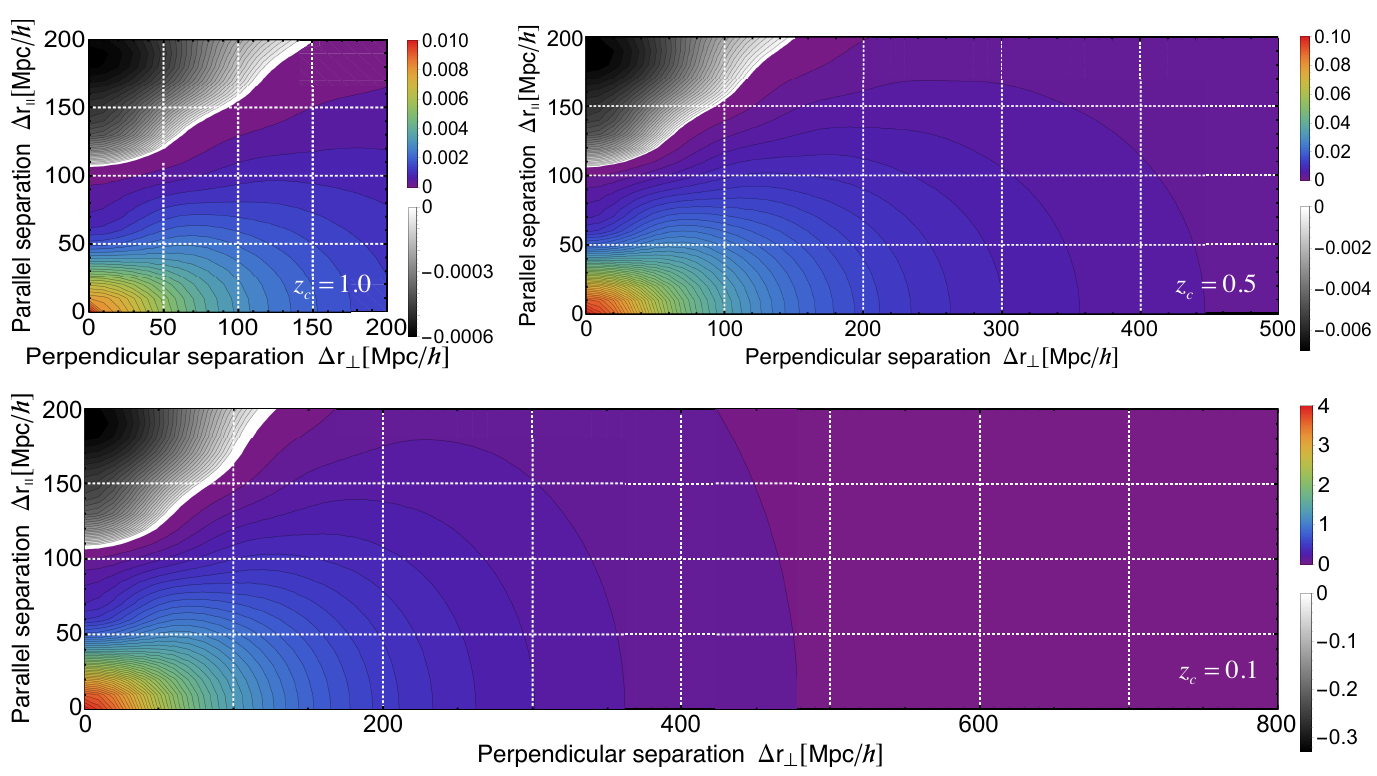}
\caption{Auto-correlation of the source velocity contribution  in the same format as Fig.~\ref{Fig:vel_obs}.}
\label{Fig:vel_source}
\end{figure}
\begin{figure}[t]
\includegraphics[height=0.5\textwidth,width=1\textwidth]{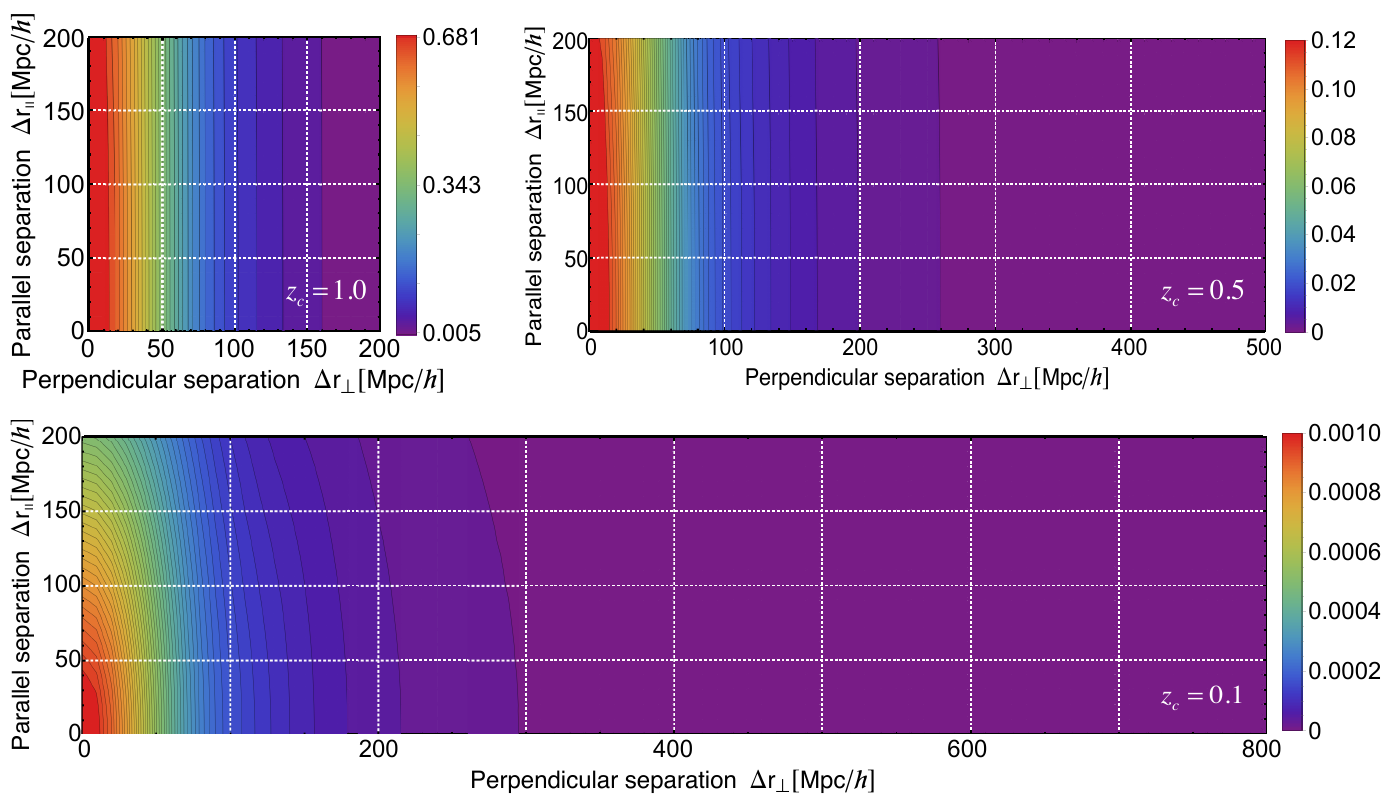}
\caption{Auto-correlation of the lensing contribution  in the same format as Fig.~\ref{Fig:vel_obs}.}
\label{Fig:lensing}
\end{figure}
\begin{figure}[t]
\includegraphics[height=0.5\textwidth,width=1\textwidth]{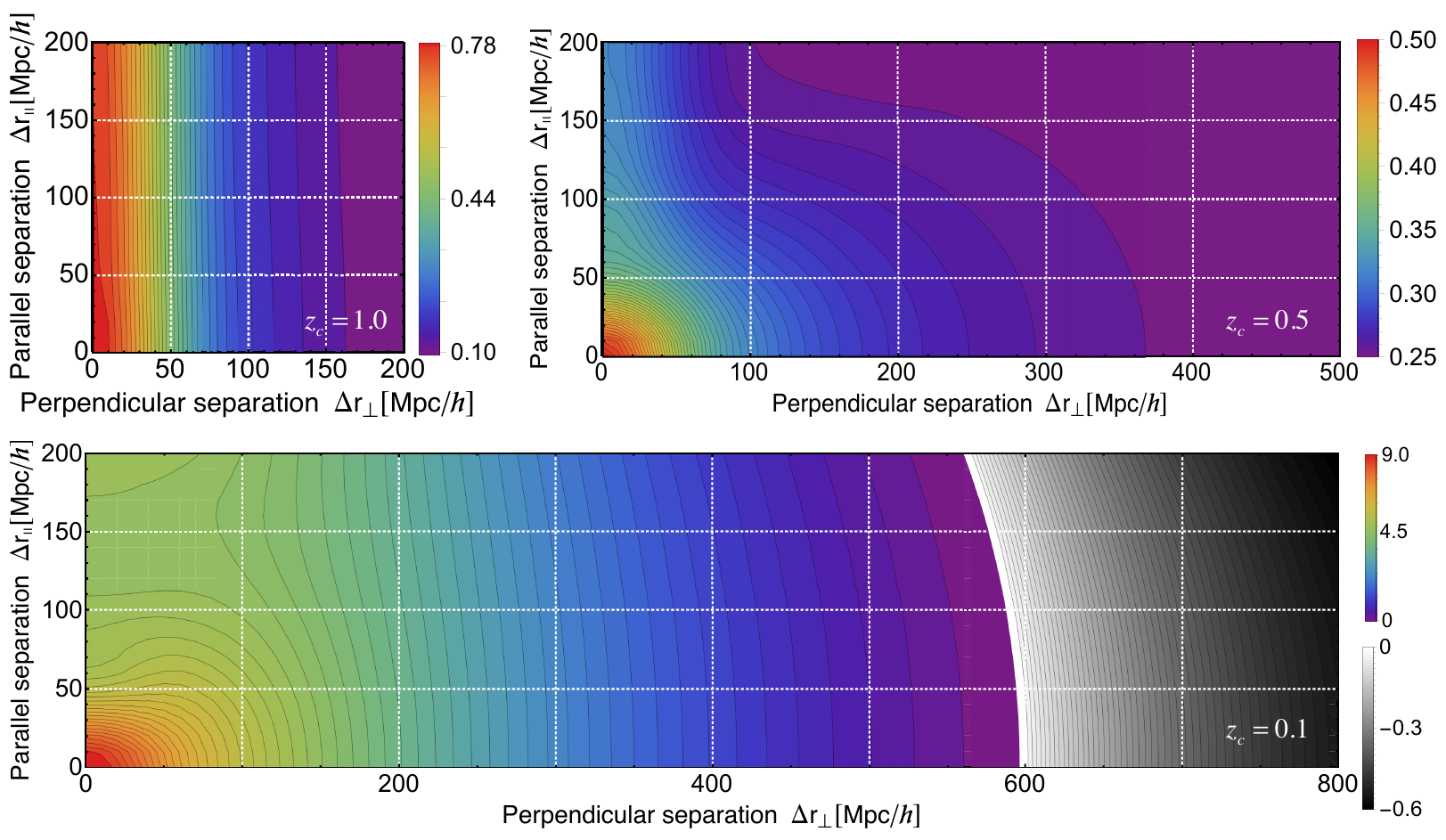}
\caption{Correlation function of the luminosity distance in the same format as Fig.~\ref{Fig:vel_obs}. The (total) correlation function is the sum of all auto-correlations in \Cref{Fig:vel_obs,Fig:vel_source,Fig:lensing} and their cross-correlations.}
\label{Fig:total}
\end{figure}

%
Figures \ref{Fig:vel_obs}, \ref{Fig:vel_source}, and \ref{Fig:lensing} show the results of the velocity at the observation, the source velocity, the lensing contributions. The sum of all the correlation functions with different midpoint redshifts are illustrated in Fig. \ref{Fig:total}. The white line indicates the contour, at which correlation vanishes. The gray color is used to represent negative values i.e., anti-correlated region.  When the magnitude in a figure changes dramatically, we make the figure with the rainbow color. In the opposite case, we use unicolor to depict subtle changes. Each figure consists of three sub-figures, and the midpoint redshifts of the sub-figures are different. Among these sub-figures, the ranges of the parallel separation are same but that of the perpendicular separations are different. The reason is that the angular coverage is usually large for small redshift  surveys.

The correlation function of the observer velocity contribution is depicted in Fig. \ref{Fig:vel_obs}. As discussed in Eq. (\ref{Eq:velocity}), the correlation of the observer velocity contribution vanishes when the angular separation between sources $\Delta\theta$ is $\Delta\theta=90^\circ$, and it becomes anti-correlated  beyond this angular separation. Since the  auto-correlation function of the observer velocity does not depend on the comoving separation between sources, the position dependency of the observer velocity contribution only comes from the coefficient of the observer velocity $(\mathcal H_z\bar r_z)^{-1}$ in Eq. (\ref{Eq:components}). In other words, the observer velocity contribution has the larger signal as the source is closer regardless of the separation between  sources. As a result, the largest signal in Fig. \ref{Fig:vel_obs} manifests at $(\Delta r_\perp,\Delta r_\| )=(0,200)$ where $(\bar r_{z_1},\bar r_{z_2})=( 193 , 393),~( 1223,1423 ),~(2214 , 2414)$ for $\bar z_c=0.1,~0.5,~1.0$. The auto-correlation becomes smaller in general as the redshift of midpoint increases. For redshift $\bar z_c=0.5$ and $\bar z_c=1.0$, the sources' comoving distances from the observer are much larger than the parallel and the perpendicular separations. Thus, the amplitude  changes very little with respect to the separations. For this reason, the plots of the observer velocity auto-correlation function of $\bar z_c=0.5$ and $\bar z_c=1.0$ are made with unicolors.

As discussed  in Eq. (\ref{Eq:velocity}) and also in Fig. \ref{Fig:vel_corr}, the auto-correlation function of the source velocity contribution in Fig. \ref{Fig:vel_source} depends not only on the distance between source and observer but also on the separation between sources. The strong signal appears at small separation where $\xi_\|$ and $\xi_\perp$ are largest. In addition, the auto-correlation function of the source velocity contribution naturally becomes negative when $\Delta r_\|>100\text{Mpc}/h$ and the parallel component is larger than the perpendicular component.   For large perpendicular separation ($\Delta r_\perp \gtrsim 500~\text{Mpc}/h$), the signal is almost constant because of the decay of $\xi_\perp$  in Fig. \ref{Fig:vel_corr}.
%

The auto-correlation of the lensing contribution is presented in Fig. \ref{Fig:lensing}. As investigated in the angular correlation function, it sharply decays as the perpendicular distance (or the angle) increases in the vicinity of the zero separation, and it becomes almost constant when $\Delta r_\perp \gtrsim 100~\text{Mpc}/h$. By contrast, the dependency on parallel separation  is very weak because the lensing contribution arises from the fluctuations projected along the line-of-sight ($\Delta r_\| \ll \bar r_c$). However, it can weakly depend on $\Delta r_\|$ at low redshift, where $\Delta r_\|\simeq \bar r_c$.



We illustrate the total correlation function of all the contributions in Fig. \ref{Fig:total}. Analogous to the angular correlation function in Fig. \ref{Fig:angular}, the velocity contribution dominates for the small midpoint redshift ($\bar z_c=0.1$). In this case, the anti-correlated region of the source velocity contribution is compensated by the large amplitude of the observer velocity contribution, and this region becomes positively correlated in the total one. When the angular separation is larger than $90~{}^\circ$, the auto-correlation of the observer velocity contribution becomes anti-correlated, and its amplitude is larger than the other contributions. As a result, the sum of all the correlations becomes anti-correlated for $\Delta\theta > 90~^\circ$. In contrast to the result of small midpoint redshift, the total correlation function of the luminosity distance at $\bar z_c=1.0$ is strongly influenced by the lensing contribution. In this case, the source velocity contribution is fully suppressed by the others. The auto-correlation of the observer velocity contribution is almost constant and it is larger than the lensing contribution for  sufficiently large perpendicular separation ($\Delta r_\perp \gtrsim 80~\text{Mpc}/h$). As a result, the total correlation at the small perpendicular separation is governed by the lensing contribution and that at the large perpendicular separation is almost constant.  The case of the midpoint redshift $\bar z_c=0.5$ is more interesting since the three contributions are all equally important, but its largest signal is still smaller than those in the other cases.  For the case of small $\Delta r_\perp$, the elongation along $\Delta r_\|$ manifests due to the lensing contribution.

\section{Discussion}
\label{Sec:discussion}

In this paper we have computed the two-point correlation function of the luminosity distances. We decomposed the luminosity distance fluctuation into the velocity, the gravitational potential, and the lensing contributions, and we have numerically evaluated each contribution. Especially, we presented the velocity at the observer position and gravitational potential contributions to the luminosity distance correlation function for the first time.\footnote{Here we do not consider the observed peculiar velocity with respect to
the CMB rest frame, in which the observed CMB dipole vanishes.} The features of each contribution are as follows. The lensing contribution dominates over the other contributions at angular separations ($\theta\lesssim 5^\circ $) and large redshift ($z\gtrsim0.5$), while the velocity contribution is the key contribution in all the other cases where the lensing contribution is small.  In contrast, the gravitational potential contribution is always subdominant to the sum of the other correlation functions, if the proper gauge-invariant expression is used. We also investigated the dependence of the correlation function of the luminosity distances on the current matter content $\Omega_m$, and   as $\Omega_m$ increases, the amplitude grows substantially. For instance, the fractional difference between the $\Omega_m=0.3$ and the $\Omega_m=0.25$ (or $\Omega_m=0.35$) amplitudes is about $30\%-40\%$ at $z=0.5$ and $z=1.0$. Especially, not only the amplitude but also the shape of the lensing correlation is affected strongly by $\Omega_m$ value.  

The expressions of the velocity and the lensing contributions to the luminosity distance are in agreement with the previous studies (e.g. \cite{HUGR06,BODUGA06,BACBM14}). From these expressions, one can derive the correlation functions of each expression. The contributions of the velocity at the source position and of the lensing to the correlation function of the luminosity distance are presented respectively in \cite{DAVIS11} and \cite{SNMB16}, and they correspond to the results in this paper. However, the expression of the gravitational potential contribution to the luminosity distance in this paper is in disagreement with the previous research, except for~\cite{SASAK87,SCJE12,JESCHI12,YOO14a}, due to the absence of the coordinate lapse at the observation. 

As shown in~\cite{BIYO16}, the gravitational potential contribution without the coordinate lapse at the observation breaks the gauge-invariance of the luminosity distance, and this absence leads to the infrared divergence in the computation of the luminosity distance correlation function. That is, the gravitational potential contribution becomes the most dominant one when perturbations on sufficiently large scales ($k_\text{IR}\rightarrow0$) are included. However, these two problems (violation of the gauge-invariance and the manifestation of the infrared divergence) are  resolved altogether by taking into account the coordinate lapse at the observation and by using the full gauge-invariant expression for the luminosity distance. With such proper expression, we showed that  the gravitational potential contribution to the correlation function of the luminosity distance is completely negligible, even if we choose very small $k_\text{IR}$.  However, the contribution of the gravitational potential without the coordinate lapse at the observer position is very close to that of the gravitational potential with the coordinate lapse when we consider only sub-horizon perturbation modes since the contribution of the coordinate lapse at the observation is only sensitive on super-horizon perturbation modes.


In observation, the observed redshift is corrected by accounting for
the peculiar motion ($v_\text{dip}\simeq 371~ \text{km/s}$ \cite{Planck:dipole}) of the observer, which is defined as the velocity needed to transform to the CMB rest frame. Since this observed peculiar velocity is well defined and independent of our gauge choice, this observational procedure of correcting the redshift causes no theoretical problems. However, providing a theoretical description of such procedure requires a gauge-invariant expression of the observed peculiar velocity with respect to the CMB rest frame, which is beyond our current scope.  In contrast, one cannot simply remove the velocity
contribution at the observer position in any chosen gauge condition,
in order to match the observational procedure, as was done in previous
work (e.g. \cite{HUGR06,HUSHSC15,BEDUET14}). For instance, the peculiar velocity field is absent in the synchronous gauge, but the luminosity distance in the synchronous gauge is equivalent to that in the other gauges due to the gauge-invariance of the luminosity distance. Therefore, if one removes the observer velocity in the conformal Newtonian gauge, it is already apparent that the resulting outcome is gauge-dependent. Second, as shown in~\cite{BIYO16}, the luminosity distance without the velocity at the observation position violates the equivalence principle, although it should be fully respected in general relativity. Furthermore, neglecting such contribution yields substantial numerical errors.

The signal-to-noise ratio of the lensing contribution to the correlation function of type Ia supernovae is investigated in~\cite{SNMB16}. According to this study, the lensing contribution to the correlation function of the luminosity distances is detectable for the deep LSST survey, where the supernova number distribution has a peak around $z\simeq 0.5$ and the coverage is 70 $\text{deg}^2$. As discussed in this paper, the velocity contribution is also comparable to the lensing contribution when sources' redshifts are around $z\sim 0.5$, and the signal-to-noise should be enhanced, taking into account the velocity contribution. That is, the correlation function of the luminosity distances is expected to be observed in the deep LSST survey, and it can be utilized as a novel tool of estimating cosmological parameters.


The dispersion of the luminosity distance measurement originates not only from the intrinsic scatter of supernovae but also from the luminosity distance fluctuation due to the inhomogeneity of the universe. In contrast to the intrinsic scatter, the fluctuation due to the inhomogeneity yields  systematic errors to the Hubble diagram. As a result, the mean luminosity distance in an inhomogeneous universe $\left<\mathcal D_L\right>$ might somewhat differ from the luminosity distance in a homogeneous universe $\bar{\mathcal D}_L$. To study this, we need to derive the second-order expression for the luminosity distance, beyond the scope of this paper,   and the $z_\text{dip}$ effect might play an important role at this level.  In fact, this computation was already performed in a few studies~\cite{BAMARI05,BEGAET13}. However, it has never been shown that the second order expressions are gauge-invariant and consistent with the equivalence principle (or at least consistent with each other). It is noted~\cite{BIYO16} that previous studies at the linear order often do not satisfy these requirements (gauge-invariance and consistency with the equivalence principle). A proper gauge-invariant calculation at the second order will be needed to completely settle the issue and quantify the shift in the mean of the luminosity distance.




\acknowledgments
We acknowledge support by the Swiss National Science Foundation, and
J.Y. is further supported by
a Consolidator Grant of the European Research Council (ERC-2015-CoG grant
680886).

\appendix

\section{The auto-correlation function of the gravitational potential contribution}
\label{App:GP}
\noindent
The correlation function of the gravitational potential contribution can be decomposed into  auto- and cross-correlations as  $\left<\delta\mathcal D_L^\Psi(z_1,\bm{\hat n_1})\delta\mathcal D_L^\Psi(z_2,\bm{\hat n_2})\right>=\xi^\Psi_{ss}+\xi^\Psi_{so}+\xi^\Psi_{oo}+\xi^\Psi_{si}+\xi^\Psi_{oi}+\xi^\Psi_{ii}$, representing the contributions at the source ($s$) and the observer ($o$), and the line-of-sight contributions ($i$). The detailed expressions of each term are 
\begin{align}
\label{Eq:GP}
	\xi_{ss}^\Psi=&~ h_1({z_1})h_1({z_2})D_{\Psi_1}D_{\Psi_2}\xi_\zeta(|\bm{r}|) \, ,\\
	\xi_{oo}^\Psi=&\left\{\left(\left(\frac{1}{\bar r_{z_1}}-\mathcal H_o h_1(z_1)	
	\right)D_{V_o}-h_2(z_1)D_{\Psi_o}\right)\times \left(1\leftrightarrow2\right)	
	\right\}\xi_\zeta(0) \, ,\nonumber\\
	\xi_{ii}^\Psi=&~4\mathcal H_{z_1}\mathcal H_{z_2}h_2(z_1)h_2(z_2)	\int_0^{\bar r_{z_1}}d\bar r_1 \int_0^{\bar r_{z_2}}d\bar r_2\left\{ D_\Psi (\bar\tau_o-\bar r_1)D_\Psi (\bar\tau_o-\bar r_2)+f_1(\Psi',\Psi'')\right\}~\xi_\zeta\left( \left|\bar r_1\hat n_1-\bar r_2\hat n_2\right|  \right)\, ,\nonumber\\
	\xi_{so}^\Psi=&\left\{\left(\frac{1}{\bar r_{z_1}}-\mathcal H_o h_1(z_1)\right)D_{V_o}-h_2(z_1)D_{\Psi_o}\right\}h_1(z_2)D_{\Psi_2}\xi_\zeta(\bar r_{z_2})
		+\left(1\leftrightarrow2\right)\, ,\nonumber\\
\xi_{oi}^\Psi=	&~2\mathcal H_{z_2}h_2(z_2)\left\{\left(\frac{1}{\bar r_{z_1}}-\mathcal H_o h_1(z_1)\right)D_{V_o}-h_2(z_1)D_{\Psi_o}
	\right\}	\int_0^{\bar r_{z_2}}d\bar r_2\left\{ D_{\Psi}(\bar\tau_o-\bar r_2)+f_2(\Psi',\Psi'')\right\} 	\xi_\zeta(\bar r_2)\nonumber\\
	&+\left(1\leftrightarrow2\right)\, ,	\nonumber\\
\xi_{si}^\Psi=	&~2\mathcal H_{z_2}h_1(z_1)h_2(z_2)D_{\Psi_1}\int_0^{\bar r_{z_2}}d\bar r_2\left\{D_\Psi(\bar\tau_o-\bar r_2)+f_2(\Psi',\Psi'')\right\}\xi_\zeta(|\bar r_{z_1}\hat n_1-\bar r_2\hat n_2|)+\left(1\leftrightarrow2\right)\, ,\nonumber
\end{align}
where $h_1(z_i)$ and $h_2(z_i)$ are defined as $h_1(z_i)\equiv \left(\frac{1}{\mathcal H_{z_i}\bar r_{z_i}}-1\right)$ and $h_2(z_i)\equiv \frac{1}{\mathcal H_{z_i}\bar r_{z_i}}$,  $f_{1}(\Psi',\Psi'')$ and $f_{2}(\Psi',\Psi'')$ indicate the (negligible) contributions of $\Psi'$ and $\Psi''$ in Eq. (\ref{Eq:components}), and $\xi_{XX}^\Psi$ and $\xi_{XY}^\Psi$ represent respectively the auto- and the cross-correlation functions ($X,Y=s,o,i$).


\begin{thebibliography}{20}

\expandafter\ifx\csname natexlab\endcsname\relax\def\natexlab#1{#1}\fi
\expandafter\ifx\csname bibnamefont\endcsname\relax
  \def\bibnamefont#1{#1}\fi
\expandafter\ifx\csname bibfnamefont\endcsname\relax
  \def\bibfnamefont#1{#1}\fi
\expandafter\ifx\csname citenamefont\endcsname\relax
  \def\citenamefont#1{#1}\fi
\expandafter\ifx\csname url\endcsname\relax
  \def\url#1{\texttt{#1}}\fi
\expandafter\ifx\csname urlprefix\endcsname\relax\def\urlprefix{URL }\fi
\providecommand{\bibinfo}[2]{#2}
\providecommand{\eprint}[2][]{\url{#2}}





\bibitem[{\citenamefont{{Perlmutter} et~al.}(1999)}]{PEADET99}
\bibinfo{author}{\bibfnamefont{S.}~\bibnamefont{{Perlmutter}}}
  \bibnamefont{et~al.}, \bibinfo{journal}{\apj} \textbf{\bibinfo{volume}{517}},
  \bibinfo{pages}{565} (\bibinfo{year}{1999}), \eprint{arXiv:astro-ph/9812133}.

\bibitem[{\citenamefont{{Riess} et~al.}(1998)}]{RIFIET98}
\bibinfo{author}{\bibfnamefont{A.~G.} \bibnamefont{{Riess}}}
  \bibnamefont{et~al.}, \bibinfo{journal}{\aj} \textbf{\bibinfo{volume}{116}},
  \bibinfo{pages}{1009} (\bibinfo{year}{1998}),
  \eprint{arXiv:astro-ph/9805201}.



\bibitem[{\citenamefont{{Ade} et~al.}(2014)}]{PLANCK13}
\bibinfo{author}{\bibfnamefont{P.~A.~R.} \bibnamefont{{Ade}}}
  \bibnamefont{et~al.}, \bibinfo{journal}{\aap} \textbf{\bibinfo{volume}{571}},
  \bibinfo{eid}{A16} (\bibinfo{year}{2014}), \eprint{arXiv:1303.5076}.


\bibitem[{\citenamefont{{Eisenstein} et~al.}(2005)\citenamefont{{Eisenstein},
  {Zehavi}, {Hogg}, {Scoccimarro} et~al.}}]{EIZEET05}
\bibinfo{author}{\bibfnamefont{D.~J.} \bibnamefont{{Eisenstein}}},
  \bibinfo{author}{\bibfnamefont{I.}~\bibnamefont{{Zehavi}}},
  \bibinfo{author}{\bibfnamefont{D.~W.} \bibnamefont{{Hogg}}},
  \bibinfo{author}{\bibfnamefont{R.}~\bibnamefont{{Scoccimarro}}},
  \bibnamefont{et~al.}, \bibinfo{journal}{\apj} \textbf{\bibinfo{volume}{633}},
  \bibinfo{pages}{560} (\bibinfo{year}{2005}), \eprint{astro-ph/0501171}.


  \bibitem[{\citenamefont{Abell}(2009)}]{LSST}
\bibinfo{author}{\bibfnamefont{P.~A.}~\bibnamefont{{Abell}} \bibnamefont{et al. (LSST)}}, \eprint{ arXiv:0912.0201}.


 
 
\bibitem[{\citenamefont{{Sasaki}}(1987)}]{SASAK87}
\bibinfo{author}{\bibfnamefont{M.}~\bibnamefont{{Sasaki}}},
  \bibinfo{journal}{\mnras} \textbf{\bibinfo{volume}{228}},
  \bibinfo{pages}{653} (\bibinfo{year}{1987}).
  
  
  
  
\bibitem[{\citenamefont{{Yoo}}(2014{\natexlab{a}})}]{YOO14a}
\bibinfo{author}{\bibfnamefont{J.}~\bibnamefont{{Yoo}}},
  \bibinfo{journal}{\cqg} \textbf{\bibinfo{volume}{31}}, \bibinfo{eid}{234001}
  (\bibinfo{year}{2014}{\natexlab{a}}), \eprint{arXiv:1409.3223}.

\bibitem[{\citenamefont{{Jeong} et~al.}(2012)\citenamefont{{Jeong}, {Schmidt},
  and {Hirata}}}]{JESCHI12}
\bibinfo{author}{\bibfnamefont{D.}~\bibnamefont{{Jeong}}},
  \bibinfo{author}{\bibfnamefont{F.}~\bibnamefont{{Schmidt}}},
  \bibnamefont{and} \bibinfo{author}{\bibfnamefont{C.~M.}
  \bibnamefont{{Hirata}}}, \bibinfo{journal}{\prd}
  \textbf{\bibinfo{volume}{85}}, \bibinfo{pages}{023504}
  (\bibinfo{year}{2012}), \eprint{arXiv:1107.5427}.

\bibitem[{\citenamefont{{Schmidt} and {Jeong}}(2016)}]{SCJE12}
\bibinfo{author}{\bibfnamefont{F.} \bibnamefont{{Schmidt}}} \bibnamefont{and}
  \bibinfo{author}{\bibfnamefont{D.}~\bibnamefont{{Jeong}}}, \bibinfo{journal}{\prd} \textbf{\bibinfo{volume}{86}},
  \bibinfo{pages}{083527 } (\bibinfo{year}{2012}), \eprint{arXiv:1204.3625}.
  
  
  
  
  \bibitem[{\citenamefont{{Bonvin} et~al.}(2006)\citenamefont{{Bonvin}, {Durrer},
  and {Gasparini}}}]{BODUGA06}
\bibinfo{author}{\bibfnamefont{C.}~\bibnamefont{{Bonvin}}},
  \bibinfo{author}{\bibfnamefont{R.}~\bibnamefont{{Durrer}}}, \bibnamefont{and}
  \bibinfo{author}{\bibfnamefont{M.~A.} \bibnamefont{{Gasparini}}},
  \bibinfo{journal}{\prd} \textbf{\bibinfo{volume}{73}},
  \bibinfo{pages}{023523} (\bibinfo{year}{2006}), \eprint{arXiv:0511183}.
  
    \bibitem[{\citenamefont{{Bacon} et~al.}(2014)\citenamefont{{Bacon}, {Andrianomena}, {Clarkson}, {Bolejko}, and {Maartens} }}]{BACBM14}
\bibinfo{author}{\bibfnamefont{D.~J.}~\bibnamefont{{Bacon}}},
  \bibinfo{author}{\bibfnamefont{S.}~\bibnamefont{{Andrianomena}}},  \bibinfo{author}{\bibfnamefont{C.}~\bibnamefont{{Clarkson}}}, \bibinfo{author}{\bibfnamefont{K.}~\bibnamefont{{Bolejko}}}, \bibnamefont{and}
  \bibinfo{author}{\bibfnamefont{R.} \bibnamefont{{Maartens}}},
  \bibinfo{journal}{\mnras} \textbf{\bibinfo{volume}{443}},
  \bibinfo{pages}{1900} (\bibinfo{year}{2014}), \eprint{arXiv:1401.3694}.
  
  




\bibitem[{\citenamefont{{Ben-Dayan}
  et~al.}(2012{\natexlab{a}})\citenamefont{{Ben-Dayan}, {Gasperini}, {Marozzi},
  {Nugier}, and {Veneziano}}}]{BEGAET12a}
\bibinfo{author}{\bibfnamefont{I.}~\bibnamefont{{Ben-Dayan}}},
  \bibinfo{author}{\bibfnamefont{M.}~\bibnamefont{{Gasperini}}},
  \bibinfo{author}{\bibfnamefont{G.}~\bibnamefont{{Marozzi}}},
  \bibinfo{author}{\bibfnamefont{F.}~\bibnamefont{{Nugier}}}, \bibnamefont{and}
  \bibinfo{author}{\bibfnamefont{G.}~\bibnamefont{{Veneziano}}},
  \bibinfo{journal}{\jcap} \textbf{\bibinfo{volume}{4}}, \bibinfo{eid}{036}
  (\bibinfo{year}{2012}{\natexlab{a}}), \eprint{arXiv:1202.1247}.

\bibitem[{\citenamefont{{Ben-Dayan}
  et~al.}(2012{\natexlab{b}})\citenamefont{{Ben-Dayan}, {Marozzi}, {Nugier},
  and {Veneziano}}}]{BEMAET12a}
\bibinfo{author}{\bibfnamefont{I.}~\bibnamefont{{Ben-Dayan}}},
  \bibinfo{author}{\bibfnamefont{G.}~\bibnamefont{{Marozzi}}},
  \bibinfo{author}{\bibfnamefont{F.}~\bibnamefont{{Nugier}}}, \bibnamefont{and}
  \bibinfo{author}{\bibfnamefont{G.}~\bibnamefont{{Veneziano}}},
  \bibinfo{journal}{\jcap} \textbf{\bibinfo{volume}{11}}, \bibinfo{eid}{045}
  (\bibinfo{year}{2012}{\natexlab{b}}), \eprint{arXiv:1209.4326}.

\bibitem[{\citenamefont{{Ben-Dayan} et~al.}(2013)\citenamefont{{Ben-Dayan},
  {Gasperini}, {Marozzi}, {Nugier}, and {Veneziano}}}]{BEGAET13}
\bibinfo{author}{\bibfnamefont{I.}~\bibnamefont{{Ben-Dayan}}},
  \bibinfo{author}{\bibfnamefont{M.}~\bibnamefont{{Gasperini}}},
  \bibinfo{author}{\bibfnamefont{G.}~\bibnamefont{{Marozzi}}},
  \bibinfo{author}{\bibfnamefont{F.}~\bibnamefont{{Nugier}}}, \bibnamefont{and}
  \bibinfo{author}{\bibfnamefont{G.}~\bibnamefont{{Veneziano}}},
  \bibinfo{journal}{\jcap} \textbf{\bibinfo{volume}{6}}, \bibinfo{eid}{002}
  (\bibinfo{year}{2013}), \eprint{arXiv:1302.0740}.

\bibitem[{\citenamefont{{Ben-Dayan} et~al.}(2014)\citenamefont{{Ben-Dayan},
  {Durrer}, {Marozzi}, and {Schwarz}}}]{BEDUET14}
\bibinfo{author}{\bibfnamefont{I.}~\bibnamefont{{Ben-Dayan}}},
  \bibinfo{author}{\bibfnamefont{R.}~\bibnamefont{{Durrer}}},
  \bibinfo{author}{\bibfnamefont{G.}~\bibnamefont{{Marozzi}}},
  \bibnamefont{and} \bibinfo{author}{\bibfnamefont{D.~J.}
  \bibnamefont{{Schwarz}}}, \bibinfo{journal}{Physical Review Letters}
  \textbf{\bibinfo{volume}{112}}, \bibinfo{pages}{221301}
  (\bibinfo{year}{2014}), \eprint{arXiv:1401.7973}.

\bibitem[{\citenamefont{{Fanizza} et~al.}(2013)\citenamefont{{Fanizza},
  {Gasperini}, {Marozzi}, and {Veneziano}}}]{FAGAET13}
\bibinfo{author}{\bibfnamefont{G.}~\bibnamefont{{Fanizza}}},
  \bibinfo{author}{\bibfnamefont{M.}~\bibnamefont{{Gasperini}}},
  \bibinfo{author}{\bibfnamefont{G.}~\bibnamefont{{Marozzi}}},
  \bibnamefont{and}
  \bibinfo{author}{\bibfnamefont{G.}~\bibnamefont{{Veneziano}}},
  \bibinfo{journal}{\jcap} \textbf{\bibinfo{volume}{11}}, \bibinfo{eid}{019}
  (\bibinfo{year}{2013}), \eprint{arXiv:1308.4935}.

\bibitem[{\citenamefont{{Gasperini} et~al.}(2011)\citenamefont{{Gasperini},
  {Marozzi}, {Nugier}, and {Veneziano}}}]{GAMAET11}
\bibinfo{author}{\bibfnamefont{M.}~\bibnamefont{{Gasperini}}},
  \bibinfo{author}{\bibfnamefont{G.}~\bibnamefont{{Marozzi}}},
  \bibinfo{author}{\bibfnamefont{F.}~\bibnamefont{{Nugier}}}, \bibnamefont{and}
  \bibinfo{author}{\bibfnamefont{G.}~\bibnamefont{{Veneziano}}},
  \bibinfo{journal}{\jcap} \textbf{\bibinfo{volume}{7}}, \bibinfo{eid}{008}
  (\bibinfo{year}{2011}), \eprint{arXiv:1104.1167}.








\bibitem[{\citenamefont{{Biern} and {Yoo}}(2016)}]{BIYO16}
\bibinfo{author}{\bibfnamefont{S.~G.} \bibnamefont{{Biern}}} \bibnamefont{and}
  \bibinfo{author}{\bibfnamefont{J.}~\bibnamefont{{Yoo}}}, \bibinfo{journal}{\jcap} \textbf{\bibinfo{volume}{1704}},
  \bibinfo{pages}{045 } (\bibinfo{year}{2017}), \eprint{arXiv:1606.01910}.


\bibitem[{\citenamefont{{Yoo} and {Gong}}(2016)}]{YOGO16}
\bibinfo{author}{\bibfnamefont{J.} \bibnamefont{{Yoo}}} \bibnamefont{and}
  \bibinfo{author}{\bibfnamefont{J.~O.}~\bibnamefont{{Gong}}}, \bibinfo{journal}{\jcap} \textbf{\bibinfo{volume}{7}},
  \bibinfo{pages}{017 } (\bibinfo{year}{2016}),
  \eprint{arXiv:1602.06300}.


\bibitem[{\citenamefont{{Yoo} and {Scaccabarozzi}}(2016)}]{YOSC16}
\bibinfo{author}{\bibfnamefont{J.} \bibnamefont{{Yoo}}} \bibnamefont{and}
  \bibinfo{author}{\bibfnamefont{F.}~\bibnamefont{{Scaccabarozzi}}}, \bibinfo{journal}{\jcap} \textbf{\bibinfo{volume}{1609}},
  \bibinfo{pages}{046 } (\bibinfo{year}{2016}),
  \eprint{arXiv:1606.08453}.





\bibitem[{\citenamefont{{Hui} and {Greene}}(2006)}]{HUGR06}
\bibinfo{author}{\bibfnamefont{L.} \bibnamefont{{Hui}}} \bibnamefont{and}
  \bibinfo{author}{\bibfnamefont{P.~B.}~\bibnamefont{{Greene}}}, \bibinfo{journal}{\prd} \textbf{\bibinfo{volume}{73}},
  \bibinfo{pages}{123526 } (\bibinfo{year}{2006}), \eprint{ astro-ph/0512159}.


\bibitem[{\citenamefont{{Huterer}, {Shafer} and {Schmidt}}(2015)}]{HUSHSC15}
\bibinfo{author}{\bibfnamefont{D.~L.} \bibnamefont{{Huterer}}}, \bibinfo{author}{\bibfnamefont{D.} \bibnamefont{{Shafer}}} \bibnamefont{and}
  \bibinfo{author}{\bibfnamefont{F.}~\bibnamefont{{Schmidt}}}, \bibinfo{journal}{\jcap} \textbf{\bibinfo{volume}{12}},
  \bibinfo{pages}{033 } (\bibinfo{year}{2015}), \eprint{ arXiv:1509.04708  }.


  
  
  
  







    \bibitem[{\citenamefont{Scovacricchi}, \citenamefont{Nichol},\citenamefont{Macaulay}, and \citenamefont{Bacon}(2016)}]{SNMB16}
\bibinfo{author}{\bibfnamefont{D.} \bibnamefont{{Scovacricchi}}}, \bibinfo{author}{\bibfnamefont{R.~C.} \bibnamefont{{Nichol}}}, \bibinfo{author}{\bibfnamefont{E.} \bibnamefont{{Macaulay}}},  \bibnamefont{and}
  \bibinfo{author}{\bibfnamefont{D.}~\bibnamefont{{Bacon}}}, \eprint{ arXiv:1611.01315 }.



 \bibitem[{\citenamefont{Cooray}, \citenamefont{Holz}, and \citenamefont{Huterer}(2016)}]{COHOHU06}
\bibinfo{author}{\bibfnamefont{A.} \bibnamefont{{Cooray}}}, \bibinfo{author}{\bibfnamefont{D.~E.} \bibnamefont{{Holz}}},  \bibnamefont{and}
  \bibinfo{author}{\bibfnamefont{D.}~\bibnamefont{{Huterer}}},  \bibinfo{journal}{\apj}
  \textbf{\bibinfo{volume}{637}}, \bibinfo{pages}{L77}
  (\bibinfo{year}{2006}), \eprint{  astro-ph/0509579 }.



    \bibitem[{\citenamefont{Aghanim}(2014)}]{Planck:dipole}
\bibinfo{author}{\bibfnamefont{N.}~\bibnamefont{{Aghanim}} \bibnamefont{et al. (Planck)}}, \bibinfo{journal}{\aap}
  \textbf{\bibinfo{volume}{571}}, \bibinfo{pages}{A27}
  (\bibinfo{year}{2014}), \eprint{ arXiv:1303.5087 }.




  

    \bibitem[{\citenamefont{Kaiser}, and \citenamefont{Hudson} (2014)}]{KAHU14}
\bibinfo{author}{\bibfnamefont{N.}~\bibnamefont{{Kaiser}, and \bibnamefont{J.} \bibnamefont{Hudson}}}, \bibinfo{journal}{\mnras}
  \textbf{\bibinfo{volume}{450}}, \bibinfo{pages}{883-895},
  (\bibinfo{year}{2014}),  \eprint{ arXiv:1411.6339}.
  

\bibitem[{\citenamefont{{Barausse} et~al.}(2005)\citenamefont{{Barausse},
  {Matarrese}, and {Riotto}}}]{BAMARI05}
\bibinfo{author}{\bibfnamefont{E.}~\bibnamefont{{Barausse}}},
  \bibinfo{author}{\bibfnamefont{S.}~\bibnamefont{{Matarrese}}},
  \bibnamefont{and} \bibinfo{author}{\bibfnamefont{A.}~\bibnamefont{{Riotto}}},
  \bibinfo{journal}{\prd} \textbf{\bibinfo{volume}{71}}, \bibinfo{eid}{063537}
  (\bibinfo{year}{2005}), \eprint{astro-ph/0501152}.



    \bibitem[{\citenamefont{Simon} (2006)}]{SI06} 
\bibinfo{author}{\bibfnamefont{P.}~\bibnamefont{{Simon}}, \bibinfo{journal}{\aap}
  \textbf{\bibinfo{volume}{473}}}, \bibinfo{pages}{711},
  (\bibinfo{year}{2007}),  \eprint{ astro-ph/0609165}.
  






    \bibitem[{\citenamefont{Yamamoto}, \citenamefont{Nakamichi}, \citenamefont{Kamino}, \citenamefont{Bassett}, and \citenamefont{Nishioka}  (2006)}]{YNKBN06}
\bibinfo{author}{\bibfnamefont{K.}~\bibnamefont{{Yamamoto}},  \bibnamefont{M.} \bibnamefont{Nakamichi}, \bibnamefont{M.} \bibnamefont{Kamino}, \bibnamefont{B.~A.} \bibnamefont{Bassett}, and \bibnamefont{H.} \bibnamefont{Nishioka}}, \bibinfo{journal}{PASJ}, \textbf{\bibinfo{volume}{58}}, \bibinfo{pages}{93},
  (\bibinfo{year}{2006}),  \eprint{ astro-ph/0505115}.

 \bibitem[{\citenamefont{Samushia}, \citenamefont{Branchini},   and \citenamefont{Percival}  (2015)}]{SABRPE15}
\bibinfo{author}{\bibfnamefont{L.}~\bibnamefont{Samushia},  \bibnamefont{E.} \bibnamefont{Branchini},  and \bibnamefont{W.~J.} \bibnamefont{Percival}}, \bibinfo{journal}{\mnras}, \textbf{\bibinfo{volume}{452}}, \bibinfo{pages}{4},
  (\bibinfo{year}{2015}),  \eprint{ arXiv:1504.02135}.



  \bibitem[{\citenamefont{Bianchi}, \citenamefont{Gil-Marín}, \citenamefont{Ruggeri},  and \citenamefont{Percival}  (2015)}]{BGRP15}
\bibinfo{author}{\bibfnamefont{D.}~\bibnamefont{{Bianchi}},  \bibnamefont{H.} \bibnamefont{Gil-Marín}, \bibnamefont{R.} \bibnamefont{Ruggeri}, and \bibnamefont{W.~J.} \bibnamefont{Percival}}, \bibinfo{journal}{\mnras}, \textbf{\bibinfo{volume}{453}}, \bibinfo{pages}{1}, (\bibinfo{year}{2015}),  \eprint{ arXiv:1505.05341}.



	


\bibitem[{\citenamefont{Davis}(2011)}]{DAVIS11} \bibinfo{author}{\bibfnamefont{T.~M.}~\bibnamefont{{Davis}} \bibnamefont{et al.}}, \bibinfo{journal}{\apj} \textbf{\bibinfo{volume}{741}}, \bibinfo{pages}{67}  (\bibinfo{year}{2011}), \eprint{ arXiv:1012.2912 }.






\bibitem[{\citenamefont{Macaulay}(2016)}]{MAC16} \bibinfo{author}{\bibfnamefont{E.}~\bibnamefont{{Macaulay}} \bibnamefont{et al.}}, \eprint{ arXiv:1607.03966 }.















 


  \bibitem[{\citenamefont{Ben-Dayan},  and \citenamefont{Takahashi}  (2015)}]{BERY2015}
\bibinfo{author}{\bibfnamefont{I.}~\bibnamefont{{Ben-Dayan}}, and \bibnamefont{R.} \bibnamefont{Takahashi}}, \bibinfo{journal}{\mnras}, \textbf{\bibinfo{volume}{455}}, \bibinfo{pages}{552}, (\bibinfo{year}{2016}),  \eprint{ arXiv:1504.07273}.


 


 \bibitem[{\citenamefont{Ben-Dayan},  and \citenamefont{Kalaydzhyan}  (2013)}]{BEKA14}
\bibinfo{author}{\bibfnamefont{I.}~\bibnamefont{{Ben-Dayan}}, and \bibnamefont{T.} \bibnamefont{Kalaydzhyan}}, \bibinfo{journal}{\prd}, \textbf{\bibinfo{volume}{90}}, \bibinfo{pages}{083509}, (\bibinfo{year}{2014}),  \eprint{ arXiv:1309.4771}



\bibitem{BE14} 
  I.~Ben-Dayan,
  arXiv:1408.3004 [astro-ph.CO].



\end{thebibliography}
\end{document}